\documentclass[preprint]{aastex}
\usepackage{lscape}

\newcommand{\sneia}{SNe\,Ia\,}
\shorttitle{$z\gtrsim1$ SNe\,Ia Host Galaxies} 
\shortauthors{Thomson and Chary}

\begin{document}

\title{Spectral Energy Distribution of $z\gtrsim1$ Type Ia Supernova Hosts in GOODS: Constraints on Evolutionary Delay and the
Initial Mass Function}
\author{M. G. Thomson$^{1}$, R. R. Chary$^2$}

\affil{$^1$ Astronomy Centre, University of Sussex, Brighton, BN1
  9QH, UK}
\affil{$^2$ Spitzer Science Center, California Institute of Technology,  Pasadena, CA 91125}
\email{m.g.thomson@sussex.ac.uk}

\begin{abstract}

We identify a sample of 22 host galaxies of Type Ia Supernovae (SNe\,Ia) at redshifts $0.95<z<1.8$ discovered in the \emph{Hubble Space Telescope} (HST) observations of the Great Observatories Origins Deep Survey (GOODS) fields. We 
measure the photometry of the hosts in 
\emph{Spitzer Space Telescope} and ground-based imaging of the GOODS fields
to provide flux densities 
from the \emph{U}-band to 24\,$\mu$m. We fit the broad-band photometry of each host with Simple Stellar Population (SSP) models to estimate the age of the stellar population giving rise to the SN\,Ia explosions. We break the well-known age-extinction degeneracy in such analyses using the \emph{Spitzer} $24\mu$m data to place upper limits on the thermally reprocessed, far-infrared emission from dust. The ages of these stellar populations give us an estimate of the delay times between the first epoch of star-formation in the galaxies
and the explosion of the SNe\,Ia. We find a bi-modal distribution of delay times ranging from 0.06 - 4.75 Gyrs. We also constrain the first-epoch of low mass star formation using these results, showing that stars of mass $\lesssim8$M$_\odot$ were formed within 3 Gyrs after the Big Bang and possibly by $z\sim6$. This argues against a truncated stellar initial 
mass function in high redshift galaxies.
\end{abstract}


\section{Introduction}

It has been shown that \sneia are standardizable
candles in that there is a very tight relation between the peak
luminosity and the width of the light-curve \citep{phillips93}. For this reason SNe\,Ia
have been used to measure the cosmic distance scale and to probe the expansion properties of the universe,
leading to the discovery that its expansion rate is
accelerating due to the existence of a repulsive force, given the name
Dark Energy \citep{riess98,perlmutter99}. It is
important to fully understand how SNe\,Ia are produced in order to minimize the systematic
uncertainties associated with these measurements and enable the properties of the Dark Energy to be investigated
as a function of cosmic time.

It is
widely believed that SNe\,Ia are the explosion of a White
Dwarf (WD) star that has reached a critical mass 
\citep[the Chandrasekhar mass,][]{chandrasekhar}. The physical mechanism leading to the WD reaching this mass is
still highly debated. Broadly, there are two leading ideas for SN\,Ia progenitors. The first is
that the WD star gains mass via accretion of material from a normal
companion star which has filled its Roche Lobe, referred to as the Single Degenerate (SD) 
scenario. The second, the Double Degenerate (DD)  scenario involves
the merger of two WD stars after formation and ejection of a common envelope in a binary system \citep[see][for a review]{livio01,podsiadlowski08}. 

In order to try to differentiate between these possible scenarios it is important
to constrain the SN\,Ia delay time, i.e. the time between the formation of the stellar
system and the supernova explosion. 
Evidence for a range of delay times has existed for some time. For example, observations show that SNe\,Ia are preferentially found in late type rather than early type galaxies, suggesting they are associated with young stellar populations with ages of 
$\sim 50$ Myrs \citep{vandenbergh90,mannucci05}. \citet{wang97} showed that SNe\,Ia are more likely to be found in the disk of a galaxy rather than the bulge, indicating an association with recent star formation and short delay times. Studies which calculate the delay time by convolving an assumed star 
formation history (SFH) with a delay time distribution (DTD) have suggested longer delay times of a few Gyrs \citep{galyam04, strolger04, dahlen04,barris06,dahlen08,strolger10}. However, \citet{forster06} show that these analyses depend strongly on the assumed SFH, giving large systematic errors, prompting \citet{oda08} to fit both the SFH and the DTD. However, those authors were only able to place
weak constraints on the DTD. From a spectroscopic study of the star-formation histories
of local SN\,Ia hosts \citet{gallagher05} put a lower limit on the delay time of 2 Gyrs.

\citet{mannucci06} have shown that a combination of observations at high and low redshift cannot be matched by a DTD with a single delay time, but that they are best matched by a bi-modal DTD. This hypothesis
is developed further by various groups \citep{scannapieco05,sullivan06,neill06,neill07} who model the SN\,Ia rate as a two-component distribution with a delayed component dependent on the host galaxy stellar mass and a prompt component dependent on the host galaxy star formation rate. Several authors have used this model to reproduce the observed SN\,Ia rates at $z\sim1$ \citep{aubourg08,neill07,dahlen04,botticella08}, although again these results depend on the assumed SFH. In order to avoid such an assumption, \citet{totani08} used an SED fitting technique to determine the ages of the host galaxy stellar populations to derive a DTD which is a power-law in the range 0.1 - 10 Gyrs. From a spectroscopic study, \citet{howell01}  showed 
that sub-luminous SNe\,Ia tend to come from old populations whereas over-luminous SNe\,Ia are from young populations,  suggesting different progenitor scenarios for the two populations. Furthermore, \citet{pritchet08} showed that the single degenerate scenario alone is not sufficient to explain the observed DTD. By studying the local environment of SNe\,Ia (rather than the properties of the whole of the host galaxy) \citet{raskin09} found that the average delay time of nearby prompt SNe was $\sim 0.3-0.5$ Gyrs, although their model allows $\sim30$\% of prompt SNe\,Ia to have delay times shorter than 0.1 Gyrs.

From a study of the X-ray properties of nearby ellipticals \citet{gilfanov10} suggest that the X-ray flux from these galaxies is consistent with only $\sim$5\% of SN\,Ia arising from accreting white dwarfs in elliptical galaxies - such systems are expected to produce X-ray emission for a significant time, whereas in the white dwarf merger (i.e. double degenerate) scenario, the X-ray emission is only present shortly before the explosion. They do not constrain the progenitors of SN\,Ia in late-type galaxies	 where the contribution from the single degenerate scenario could be significant. Using a maximum-likelihood inversion procedure, \citet{maoz10} recover the DTD for a sample of local SNe\,Ia utilising the SFH of each individual galaxy, finding evidence for both prompt (with delay times $<0.42$ Gyrs)  and delayed (with delay times $>2.4$ Gyrs) SNe, where the DTD has a peak at short delay times but a broad distribution to longer delay times. Similar conclusions are reached by \citet{brandt10}.

 \citet{greggio10} build on the parameterisation of \citet{greggio05} of the evolution of binary systems to calculate some features of the DTD of single-degenerate, double degenerate and mixed model progenitor scenarios. All models are consistent with both prompt and delayed SNe, although the distributions are continuous (suggesting that the distinction is arbitrary).
 They do explore scenarios where there is a mix of SD and DD and where the SD contributes more prompt SNe than the DD channel, however, they see no theoretical basis for this and comparisons to the observed SN\,Ia rates of \citet{greggio09} and \citet{sullivan06} do not favour the mixed scenarios.

The Great Observatories Origins Deep Survey \citep[GOODS;][]{dickinson03} is a multi-wavelength survey covering $\sim330$ arcmin$^2$ over two 
fields.  In conjunction with the GOODS survey a supernova search was conducted, surveying both fields at several different epochs with HST ACS \citep{giavalisco04} resulting in a catalogue of 22 SNe\,Ia at $z \ge 0.95$ with spectroscopic redshifts, \citep{riess04, riess07}. We add a further 3 SNe 1997fg, 1997ff and 2002dd from past SN searches in these fields \citep{gilliland99,riess01,blakeslee03}.

In the present work, we compile 
optical/near-infrared photometry from ACS and NICMOS on HST, \emph{Spitzer} $3.6-24\,\mu$m data
as well as supplementary ground-based data where available for the host galaxies of these SNe\,Ia.
We then fit the multi-wavelength photometry of the
hosts to the Single Stellar Population (SSP) models of Charlot and
Bruzual (priv. com.) in order to find the ages of the stellar
populations in the SN host galaxies, thus allowing us to study the delay times of SNe\,Ia.

Furthermore, by calculating the ages of the stellar populations we can constrain the first epoch of low mass star formation \citep[the progenitors of SNe\,Ia have masses $\lesssim 8\textrm{M}_\odot$, see e.g. ][and references therein]{blanc08}, i.e. the
time after the Big Bang that stars must have formed in order to yield
the stellar populations we find. This allows us to constrain models which suggest that only stars with $\gtrsim10$M$_{\sun}$ might have formed at $z\sim6$ \citep{tumlinson04}.

The remainder of this paper is arranged as follows. In section 2 we describe the identification of host galaxies, in section 3 we outline the SED fitting method. We present our results in section 4 and provide a discussion of these results in section 5. Finally our conclusions are given in section 6. In the rest of this analysis we use the concordance cosmology of $\Omega_M = 0.3$, $\Omega_\Lambda = 0.7$ and $H_0=70$ km s$^{-1}$ Mpc$^{-1}$. We present photometry in the AB magnitude system unless otherwise stated.

\section{Host Galaxy Identification}

\begin{figure}[!h]	
\centering
 \includegraphics[width=7cm]{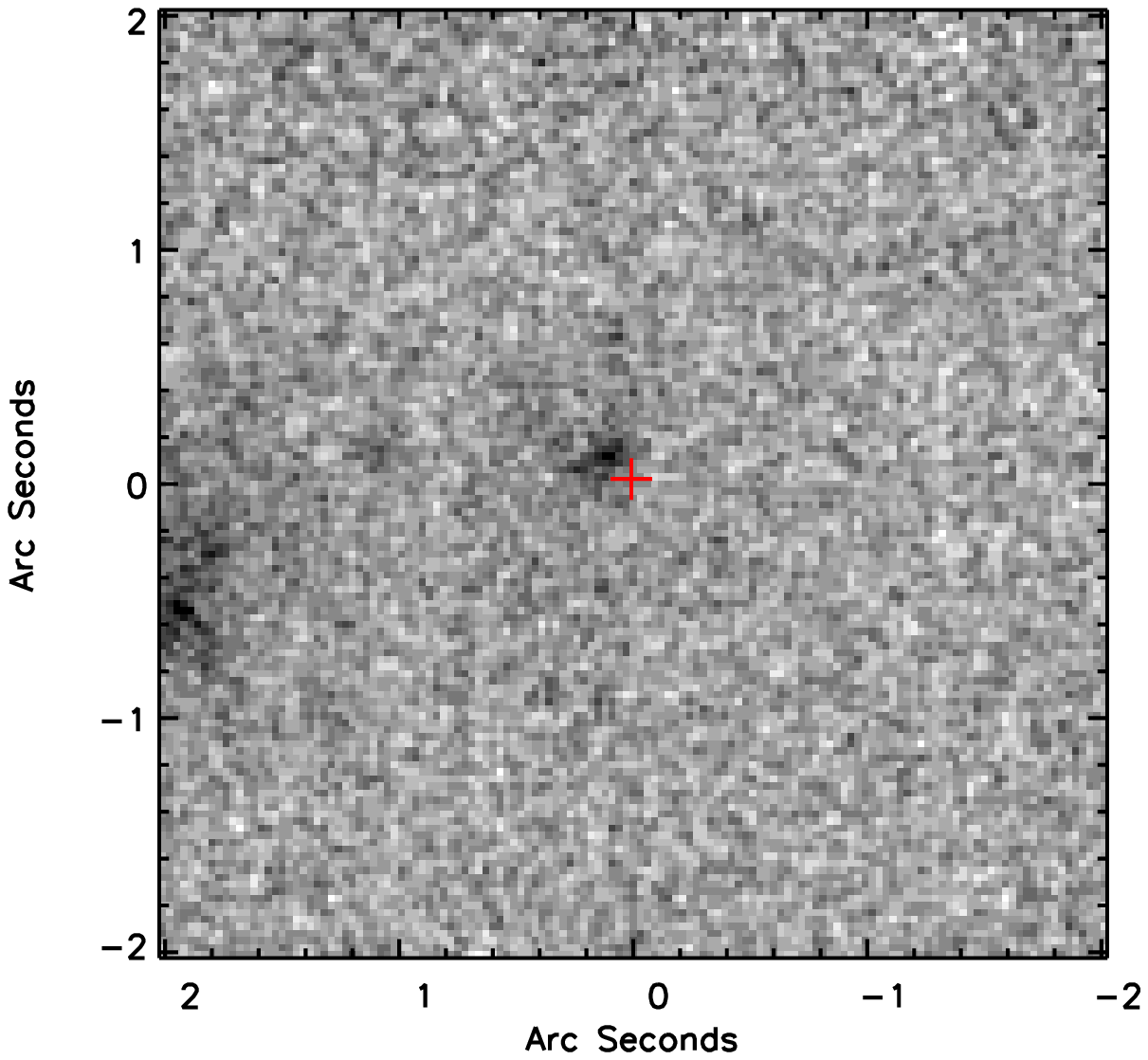}
 \includegraphics[width=7cm]{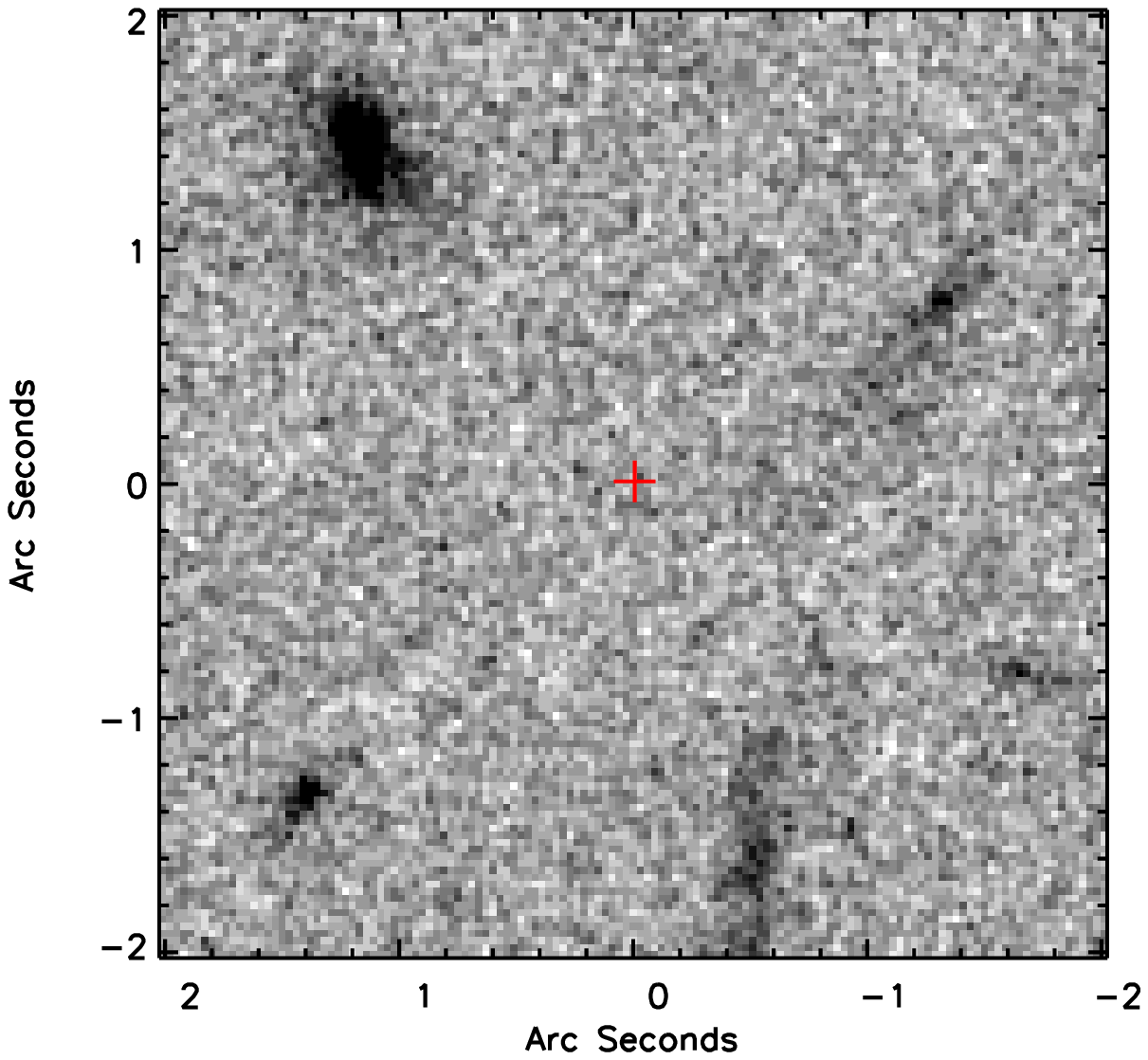}
 \caption[Host Identification Postage Stamps]{HST \emph{V} band images centred around two \sneia, the positions of which are indicated by the red cross. The left hand panel shows the host galaxy for 1997ff is well identified whereas the right hand panel shows that the host for HST04Gre has not been identified. It is therefore removed from the 
sample of host galaxies which are analysed in this work.}\label{fig:hoststamps}
 \end{figure}

We use the supernova sample of \citet{riess07} with additional SNe\,Ia from \citet{gilliland99} and \citet{blakeslee03}, giving us 25 SNe\,Ia at
$z \ge 0.95$. We cross-matched these SNe\,Ia with the GOODS ACS v2.0 catalogue
in both the Northern and Southern Fields using the software \emph{topcat}
\citep{topcat} and a cross-matching radius of 1\mbox{$^{\prime\prime}$}. This resulted
in a catalogue of 22 SNe\,Ia host galaxies. We then match
the ACS host galaxy positions to the ground-based and IRAC positions
with a radius of 0\farcs5, after correcting the IRAC and ground-based catalogues for the well-known 0\farcs38 offset in the GOODS North catalogues. The SN positions are also corrected for this offset where necessary. The large radius used for our initial
cross-match ensures that all possible SNe\,Ia with detected ACS host
galaxies are identified, however, it could lead to false
identifications with other galaxies. Therefore we visually inspected
the ACS images for each SN\,Ia, but all appeared to be good identifications. The three SNe with no matches are 2002dd, HST04Sas and HST04Gre. In the cases of 2002dd and HST04Sas, the host is not found in the catalogue due to confusion with nearby bright galaxies. HST04Gre appears to be a hostless SN\,Ia, the \emph{V}-band postage stamp is shown in figure~\ref{fig:hoststamps} along with the host for SN 1997ff in the same band, where the host is well identified. The co-ordinates of the SNe\,Ia and their host galaxies are given in Table~\ref{tab:coords} (SNe\,Ia with ambiguous host identifications are shown with only the SN coordinates given).

\begin{deluxetable}{lcccc}
\tabletypesize{\small}
\tablewidth{0pt}
\tablecolumns{6}
\tablecaption{J2000 coordinates of the SNe\,Ia and the offset to the  host galaxies along with the spectroscopic  redshift of the SNe\,Ia (either spectroscopy directly of the SN\,Ia or from spectroscopy of the host). SNe\,Ia where the host galaxy offset is missing have a confused or undetected host and are therefore dropped from the sample.\label{tab:coords}}
\tablehead{
\colhead{SN Name} & \colhead{SN RA} & \colhead{SN Dec} & \colhead{Host Offset [\arcsec]} & \colhead{redshift}} 
\startdata
1997ff & 189.18379 & 62.21244 & 0.15 & 1.75\\
2003es & 189.23079 & 62.21987 & 0.48 & 0.95\\
2003az & 189.33196 & 62.31031 & 0.16 & 1.26\\
2003dy & 189.28817 & 62.19128 & 0.29 & 1.34\\
2002ki & 189.36813 & 62.34434 & 0.33 & 1.14\\
HST04Pat & 189.53750 & 62.31312 & 0.41 & 0.97\\
HST05Fer & 189.10458 & 62.25662 & 0.66 & 1.02\\
HST05Koe & 189.09550 & 62.30644 & 0.39 & 1.23\\
HST05Red & 189.25708 & 62.20666 & 0.39 & 1.19\\
HST05Lan & 189.23633 & 62.21481 & 0.74 & 1.23\\
HST04Tha & 189.22987 & 62.21779 & 0.39 & 0.95\\
HST04Eag & 189.33705 & 62.22799 & 0.46 & 1.02\\
HST05Gab & 189.05763 & 62.20210 & 0.40 & 1.12\\
HST05Str & 189.08596 & 62.18072 & 0.05 & 1.01\\
1997fg & 189.24029 & 62.22092 & 0.39 & 0.95\\
2003aj & 53.18471 & -27.91844 & 0.20 & 1.31\\
2002fx & 53.02833 & -27.74289 & 0.26 & 1.40\\
2003ak & 53.19542 & -27.91372 & 0.36 & 1.55\\
2002hp & 53.10329 & -27.77161 & 0.19 & 1.30\\
2002fw & 53.15633 & -27.77961 & 0.51 & 1.30\\
HST04Mcg & 53.04250 & -27.83055 & 0.57 & 1.37\\
HST04Omb & 53.10558 & -27.75084 & 0.22 & 0.98\\
HST04Sas & 189.22546 & 62.13966 & - & 1.39\\
2002dd & 189.23067 & 62.21281 & - & 0.95\\
HST04Gre & 53.08954 & -27.78286 & - & 1.14\\
\enddata
\tablenotetext{a}{Typographical error in the co-ordinates given in \citet{riess07} discovered by comparison to their postage stamps. Original co-ordinates are HST04Eag  189.33646, 62.22820; HST04Mcg   53.04175,  -27.83055, corrected co-ordinates are those presented here.}
\end{deluxetable}

In order to break the well-known degeneracy that exists between age and extinction in fitting 
model spectral energy distributions to optical/NIR photometry, we use the MIPS 24$\mu$m photometry to place constraints on the fraction of energy that is absorbed by dust and re-emitted at longer wavelengths. This results in upper limits
on the amount of extinction applied to the model SED. 

For each host we then have HST ACS \emph{BViz}, \emph{Spitzer} IRAC ch$1-4$\footnote{with the exception of SNe HST04Pat, which was off the edge of the complete \emph{Spitzer} survey, and 2002ki which had an uncertain IRAC counterpart due to confusion} and MIPS 24$\mu$m$^1$ data as well as ground based\footnote{Ground-based data is from ISAAC/Very Large Telescope, Keck Telescope (spectroscopy), MOSAIC/Kitt Peak National Observatory, WIRCAM on the Canada-France Hawaii Telescope and the MOSAIC/Cerro Tololo Inter-American Observatory. The full list of data sets can be found at {\rm http://www.stsci.edu/science/goods/}} \emph{UJHK}. We also have some HST NICMOS \emph{JH} coverage from \citet{buitrago08}. We reject photometry with statistical errors larger than 0.3 mag which are indicative of marginal detections. We then add a systematic error in quadrature to the photometric errors given in the catalogue of $5\%$ for HST ACS and NICMOS \citep{acshandbook, nicmoshandbook}, $10\%$ for IRAC ch$1 - 3$ and ground-based photometry and $15\%$ for IRAC ch4 \citep{reach05}. We remove the southern \emph{U} band data from the analysis due to inhomogeneous variations across the field. 

The process of mosaicing the ACS data includes the flux from the SNe in the \citet{riess07} sample. We have estimated and subtracted out the SN contribution to the host galaxy
photometry in the catalogue using the following process. We use the SN photometry given in \citet{riess04,riess07}. In each band we add up the total SN flux and divide by the number of epochs of observations going into the ACS stack and subtract this from the host flux. This effect is present in the \emph{i} and \emph{z} bands, with a small contribution in the \emph{V}-band (partly due to the SN colour and partly because the later \citet{riess07} SN search did not include \emph{V}-band re-imaging of the GOODS fields). The \emph{B}-band data was taken at an earlier time to the \emph{Viz} data and is thus not affected. In several cases \citet{riess04} give \emph{i} and \emph{z} band SN photometry but not the \emph{V}-band photometry. In these cases, we estimate the contribution to the $V-$band by extrapolating a simple power-law spectrum fit to the nearest two bands. In a very few cases, only $z-$band photometry is given; in these cases we fit the power-law to the observed colours of Type Ia SNe as presented in \citet{jha07} to measure the contribution of the SN in the $i-$band and the $V-$band. In all cases the correction to the host photometry in the \emph{V}-band is small (there are only 3 cases where the correction is larger than 0.01$\mu$Jy with the largest being 0.076$\mu$Jy). In the \emph{i} and \emph{z}-bands the correction is small in most cases, the correction is larger than $\sim10$\% of the host flux in only 4 and 5 cases respectively.  In the cases of HST05Gab and 2002fw the SN contamination is very large, giving unreliable photometry in the \emph{i} and \emph{z} bands. Removing these bands leaves us with good photometry only in the \emph{BV}, IRAC ch1 and ch2 in the case of HST05Gab and only \emph{BV} and \emph{K} in the case of 2002fw. Since we are unlikely to be able to constrain the host SED with only 3 or 4 bands we remove these SNe from the analysis. In the remaining cases a visual inspection further suggests that the contamination is small. Therefore, although the details of the stack process could affect the precise correction that is required, this effect is likely to be small.

The final photometry used, including the systematic errors added in quadrature is given in tables~\ref{tb:photometry} (HST ACS, ground-based) and ~\ref{tb:photometry_irac} (\emph{Spitzer} IRAC and MIPS). The photometry used in the present work for the host galaxy of SN 1997ff differs from that given in \citet{riess01}. In particular, \citet{riess01} give \emph{B}-band photometry of $26.67 \pm 0.16$ mag, whereas we do not find an accurate detection for the host-galaxy in this band (the catalogue value is $28.23\pm1.26$ mag so it is removed from our analysis due to the large error). Further investigation revealed that the host is given in \citet{williams96} as \emph{B}-band magnitude of 29.18 with a signal-to-noise of 3.5. A similar pattern is seen in the \emph{iz} bands suggesting that perhaps there is a variable Active Galactic Nucleus (AGN) in this host, although further investigation beyond the scope of this work is required to confirm this. 

\begin{landscape}
\begin{deluxetable}{lcccccccc}
\tabletypesize{\small}
\tablewidth{0pt}
\tablecolumns{9}
\tablecaption{HST ACS and ground-based photometry of the host galaxies used in the fits, the errors given include additional systematic errors (see text). All magnitudes are in AB magnitudes.\label{tb:photometry}}
\tablehead{
\colhead{SN Name} & \colhead{\emph{U} [mag]} & \colhead{\emph{B} [mag]} & \colhead{\emph{V} [mag]} & \colhead{\emph{i} [mag]} & \colhead{\emph{z} [mag]} & \colhead{\emph{J} [mag]} & \colhead{\emph{H} [mag]} & \colhead{\emph{K} [mag]}}
\startdata
1997ff & - &- & 26.11$\pm$ 0.16 & 25.06$\pm$ 0.10 & 23.97$\pm$ 0.06 & 22.59$\pm$ 0.05* & 21.60$\pm$ 0.05* & - \\
2003es & - & 25.86$\pm$ 0.17 & 24.12$\pm$ 0.06 & 22.54$\pm$ 0.05 & 21.65$\pm$ 0.05 & 21.11$\pm$ 0.05* & 20.44$\pm$ 0.05* & 19.66$\pm$ 0.11 \\
2003az & - & 26.26$\pm$ 0.17 & 26.74$\pm$ 0.21 & 25.14$\pm$ 0.08 & 24.30$\pm$ 0.06 & - & - & - \\
2003dy & 23.75$\pm$ 0.11 & 23.56$\pm$ 0.06 & 23.49$\pm$ 0.05 & 23.28$\pm$ 0.06 & 22.76$\pm$ 0.05 & - & - & 21.96$\pm$ 0.24 \\
2002ki & 24.31$\pm$ 0.11 & 24.37$\pm$ 0.07 & 24.26$\pm$ 0.06 & 23.88$\pm$ 0.06 & 23.45$\pm$ 0.06 & - & - & 21.97$\pm$ 0.28 \\
HST04Pat & 22.49$\pm$ 0.10 &- & 21.58$\pm$ 0.06 & 20.80$\pm$ 0.05 & 20.39$\pm$ 0.05 & 20.24$\pm$ 0.14 & - & 19.94$\pm$ 0.14 \\
HST05Fer & 25.89$\pm$ 0.17 & 25.97$\pm$ 0.07 & 25.84$\pm$ 0.07 & 25.05$\pm$ 0.06 & 24.52$\pm$ 0.06 & - & - & - \\
HST05Koe & 26.33$\pm$ 0.23 &- & 25.02$\pm$ 0.12 & 24.09$\pm$ 0.09 & 23.45$\pm$ 0.07 & - & - & 21.80$\pm$ 0.28 \\
HST05Red & 24.76$\pm$ 0.11 & 24.62$\pm$ 0.06 & 24.61$\pm$ 0.06 & 24.34$\pm$ 0.06 & 24.14$\pm$ 0.06 & 23.87$\pm$ 0.05* & 23.72$\pm$ 0.05* & - \\
HST05Lan & - & 26.45$\pm$ 0.19 & 25.31$\pm$ 0.07 & 24.25$\pm$ 0.06 & 23.30$\pm$ 0.05 & 22.59$\pm$ 0.05* & 21.84$\pm$ 0.05* & - \\
HST04Tha & - & 26.99$\pm$ 0.22 & 25.24$\pm$ 0.06 & 23.94$\pm$ 0.05 & 23.04$\pm$ 0.05 & 22.60$\pm$ 0.05* & 21.99$\pm$ 0.05* & 21.48$\pm$ 0.14 \\
HST04Eag & 24.05$\pm$ 0.11 & 23.93$\pm$ 0.06 & 23.56$\pm$ 0.06 & 22.91$\pm$ 0.05 & 22.57$\pm$ 0.05 & 21.96$\pm$ 0.23 & - & 21.19$\pm$ 0.15 \\
HST05Gab & - & 26.78$\pm$ 0.19 & 26.76$\pm$ 0.14 &- & - & - & - & - \\
HST05Str & 24.77$\pm$ 0.11 & 24.33$\pm$ 0.07 & 24.04$\pm$ 0.06 & 23.50$\pm$ 0.06 & 23.12$\pm$ 0.06 & - & - & - \\
1997fg & 24.31$\pm$ 0.11 & 23.99$\pm$ 0.06 & 23.56$\pm$ 0.06 & 22.90$\pm$ 0.06 & 22.59$\pm$ 0.05 & 22.12$\pm$ 0.25 & - & 21.90$\pm$ 0.23 \\
2003aj & - & 25.09$\pm$ 0.07 & 24.99$\pm$ 0.06 & 24.85$\pm$ 0.07 & 24.23$\pm$ 0.06 & 23.96$\pm$ 0.15 & - & 23.67$\pm$ 0.17 \\
2002fx & - & 25.59$\pm$ 0.08 & 25.90$\pm$ 0.09 & 25.62$\pm$ 0.10 & 25.50$\pm$ 0.09 & 25.28$\pm$ 0.22 & 25.28$\pm$ 0.29 & 24.92$\pm$ 0.28 \\
2003ak & - & 23.66$\pm$ 0.06 & 23.57$\pm$ 0.06 & 23.39$\pm$ 0.06 & 23.14$\pm$ 0.06 & 22.57$\pm$ 0.11 & - & 22.45$\pm$ 0.12 \\
2002hp & - &- & 25.86$\pm$ 0.11 & 24.04$\pm$ 0.06 & 23.15$\pm$ 0.06 & 21.80$\pm$ 0.11 & 21.25$\pm$ 0.11 & 20.71$\pm$ 0.10 \\
HST04Mcg & - & 25.09$\pm$ 0.11 & 24.54$\pm$ 0.08 & 23.74$\pm$ 0.07 & 23.06$\pm$ 0.06 & 22.16$\pm$ 0.11 & 21.63$\pm$ 0.11 & 21.21$\pm$ 0.11 \\
HST04Omb & - & 23.17$\pm$ 0.05 & 23.16$\pm$ 0.05 & 22.86$\pm$ 0.05 & 22.69$\pm$ 0.05 & 22.56$\pm$ 0.11 & 22.85$\pm$ 0.12 & 22.49$\pm$ 0.12 \\
\enddata
\tablenotetext{a}{NICMOS photometry}
\end{deluxetable}
\end{landscape}

\begin{landscape}
\begin{deluxetable}{lccccc}
\tabletypesize{\small}
\tablewidth{0pt}
\tablecolumns{6}
\tablecaption{\emph{Spitzer} IRAC and MIPS photometry of the host galaxies used in the fits, the errors given include additional systematic errors (see text). \label{tb:photometry_irac}}
\tablehead{
\colhead{SN Name} & \colhead{3.6 [$\mu$Jy]}   & \colhead{4.5$\mu$m [$\mu$Jy]}   & \colhead{5.8$\mu$m [$\mu$Jy]}   & \colhead{8.0$\mu$m [$\mu$Jy]}   & \colhead{24.0$\mu$m [$\mu$Jy]}}
\startdata
1997ff & 30.1$\pm$ 3.0 & 33.3$\pm$ 3.3 & 29.4$\pm$ 3.0 & 20.0$\pm$ 3.0 & 27.5$\pm$ 4.8 \\
2003es & 56.6$\pm$ 5.7 & 39.3$\pm$ 3.9 & 29.6$\pm$ 3.0 & 20.4$\pm$ 3.1 & $<25$ \\
2003az & 20.1$\pm$ 2.0 & 20.3$\pm$ 2.0 & 15.4$\pm$ 1.6 & 10.4$\pm$ 1.6 & $<25$ \\
2003dy & 7.6$\pm$ 0.8 & 6.7$\pm$ 0.7 & 5.0$\pm$ 0.6 & 4.4$\pm$ 0.7 & $<25$ \\
2002ki & $<$ 0.5 & $<$ 0.5 & $<$ 1.5 & $<$ 1.5 & $<25$ \\
HST04Pat & - & - & $<$ 1.5 & - & $<25$ \\
HST05Fer & 2.1$\pm$ 0.2 & 1.3$\pm$ 0.1 & 1.3$\pm$ 0.4 & - & $<25$ \\
HST05Koe & 16.3$\pm$ 1.6 & 13.2$\pm$ 1.3 & 11.7$\pm$ 1.3 & 8.2$\pm$ 1.4 & 63.1$\pm$ 6.6 \\
HST05Red & 2.5$\pm$ 0.2 & 1.9$\pm$ 0.2 & - & 1.2$\pm$ 0.4 & $<25$ \\
HST05Lan & 21.9$\pm$ 2.2 & 18.6$\pm$ 1.9 & 12.2$\pm$ 1.3 & 8.8$\pm$ 1.4 & $<25$ \\
HST04Tha & 13.4$\pm$ 1.3 & 9.0$\pm$ 0.9 & 6.4$\pm$ 0.7 & 3.4$\pm$ 0.6 & $<25$ \\
HST04Eag & 10.3$\pm$ 1.0 & 7.6$\pm$ 0.8 & 5.8$\pm$ 0.7 & 5.7$\pm$ 0.9 & 50.9$\pm$ 5.1 \\
HST05Gab & 0.6$\pm$ 0.1 & 0.3$\pm$ 0.1 & $<$ 1.5 & $<$ 1.5 & $<25$ \\
HST05Str & 5.4$\pm$ 0.5 & 3.8$\pm$ 0.4 & 2.8$\pm$ 0.5 & 2.3$\pm$ 0.6 & $<25$ \\
1997fg & 7.8$\pm$ 0.8 & 5.6$\pm$ 0.6 & 4.5$\pm$ 0.5 & 3.7$\pm$ 0.6 & $<25$ \\
2003aj & 1.6$\pm$ 0.2 & 1.3$\pm$ 0.1 & - & $<$ 1.5 & $<25$ \\
2002fx & 0.4$\pm$ 0.1 & 0.4$\pm$ 0.1 & $<$ 1.5 & - & $<25$ \\
2003ak & 5.0$\pm$ 0.5 & 5.1$\pm$ 0.5 & 3.3$\pm$ 0.5 & 2.2$\pm$ 0.6 & $<25$ \\
2002hp & 35.4$\pm$ 3.5 & 31.9$\pm$ 3.2 & 22.2$\pm$ 2.2 & 15.4$\pm$ 2.3 & $<25$ \\
HST04Mcg & 20.8$\pm$ 2.1 & 20.1$\pm$ 2.0 & 14.7$\pm$ 1.5 & 13.1$\pm$ 2.0 & 59.5$\pm$ 6.0 \\
HST04Omb & 4.3$\pm$ 0.4 & 3.0$\pm$ 0.3 & 1.8$\pm$ 0.4 & - & $<25$ \\
\enddata
\end{deluxetable}
\end{landscape}


\section{Model Fitting}
We use the population synthesis models of Charlot and Bruzual
(CB07, priv. com.) which are generated from an updated version of the \citet{bruzual03} GALAXEV code to produce models
which include a new prescription for Thermally  Pulsating AGB stars. 
The code calculates the spectral evolution of a stellar population based on a library of observed stellar spectra (see \citet{bruzual03} and references therein). We generated a
suite of models with varying metallicity, Initial Mass Function (IMF; i.e. the distribution of stellar masses for the starburst), stellar population age and star formation history (SFH, either a constant SFR or an exponentially declining SFR with e-folding time $\tau$). For each model we apply a dust extinction correction of varying $A_{V}$ with a \citet{calzetti00} dust extinction law, giving a total of $\sim 5\times 10^5$ models. Table~\ref{tb:modelparams} shows the parameter values used to generate the models. Finally, we ensure that the age of the stellar population never exceeds the age of the Universe at the redshift of the SN\,Ia.

\begin{landscape}
\begin{deluxetable}{cl}
\tablecolumns{2}
\tabletypesize{\small}
\tablewidth{0pt}
\tablecaption{Input parameters used to generate the model SEDs \label{tb:modelparams}}
\tablehead{
\colhead{Parameter}  & \colhead{Allowed Values}} 
\startdata
Metallicity & 0.005, 0.020, 0.200, 0.400, 1.000, 2.500 Z$_\odot$\\[11pt]
\hline
 & Salpeter:  $\frac{\textrm{d}n}{\textrm{d}M} = M^{-1.35} \qquad 0.1\textrm{M}_\odot < M < 100\textrm{M}_\odot$\\
IMF &  \\
 & Chabrier: $\frac{\textrm{d}n}{\textrm{d}M} = \textrm{exp}(-(\log_{10}(M) - \log_{10}(0.08))^2/0.9522)\quad 0.1\textrm{M}_\odot < M < 1\textrm{M}_\odot$\\
  &                 $\qquad \qquad \qquad \quad= M^{-1.3} \qquad \qquad \qquad \qquad \qquad \qquad \quad 0.1\textrm{M}_\odot < M < 100\textrm{M}_\odot$\\[11pt]
  \hline
  SFH & $\psi(t) = \tau^{-1}\textrm{exp}(-t/\tau), \quad  \tau = $ 0.01, 0.03, 0.1, 0.3 1 Gyrs and $\tau = \infty$ i.e. constant SFR \\[11pt]
  \hline
  Stellar population age & unevenly spaced in the range 0.001 - 6 Gyrs \\[11pt]
  \hline
  Dust, $A_V$ &  0 - 5 mag, step size of 0.1 mag, starburst extinction law \citep{calzetti00}
  \enddata
\end{deluxetable}
\end{landscape}

Each model is then converted to the observed frame using the
spectroscopic redshift of the SN\,Ia and convolved with the response/filter functions of the
instruments in order to generate the equivalent observed photometry for each model in each
band. We then calculate the $\chi^2$ of each model
compared to the data, according to the equation 
\[
\chi^2 = \sum \frac{(f^o_i - bf^m_i)^2}{\sigma^2_i}
\]
where $f^o$ is the observed flux in each band, $f^m$
is the model flux in each band, $\sigma$ is the observed error
(including the additional systematic error), $b$ is a normalization
factor (calculated as the mean ratio of the observed flux to
the model flux, weighted by the errors) and the summation is over all
bands. We then find the model with the minimum $\chi^2$ for each host galaxy.

In order to break the degeneracy between young models with large extinction and old models with low extinction we use the 24$\mu$m photometry since it is a reliable proxy for the fraction of light that is absorbed by dust and reprocessed at longer wavelengths. By using the \citet{chary01} templates, the 24$\mu$m photometry can be translated to a total far-infrared (FIR) luminosity. This provides an upper limit to the fraction of optical/near-infrared light that is absorbed by dust. We then compare this to the FIR luminosity of each model, calculated thus
\[
I= \int l_\lambda (1-\textrm{exp}(-A_V \kappa/1.086)) \mathrm{d} \lambda
\]
where $l_\lambda$ is the flux of the model at each wavelength $\lambda$ and $\kappa$ is the extinction correction given by the extinction law of \citet{calzetti00}. Any models which have an FIR luminosity that is in excess of the FIR luminosity calculated from the 24$\mu$m photometry/limits are rejected.
We use upper limits to reject any models that exceed the photometric limits. The limits we use are 28.6, 28.6, 27.9, 27.4 AB mag for the ACS \emph{BViz} bands (5$\sigma$ limits), 0.5, 0.5, 1.5, 1.5$\mu$Jy for the IRAC ch1-4 and 25$\mu$Jy for the MIPS 24$\mu$m data (80\% completeness limits).

Each SED in the CB07 models is the combination of the SEDs from all the various stars that have formed  (and evolved) over the lifetime of the model galaxy. All but one of our star-formation histories is exponential, in these cases  most of the stars are formed at the redshift of formation. The final time-step from the models is the upper limit to the age of a star in the population.  While most of the stars in the population are old, most of the luminosity comes  from younger stars. In order to account for this we calculate a mean age weighted by the fractional contribution to the $V$-band luminosity from the stars of each age. 

\section{Results}

\subsection{Delay Time Distribution}
\begin{figure}[htb]
 \centering
\includegraphics[width=14cm]{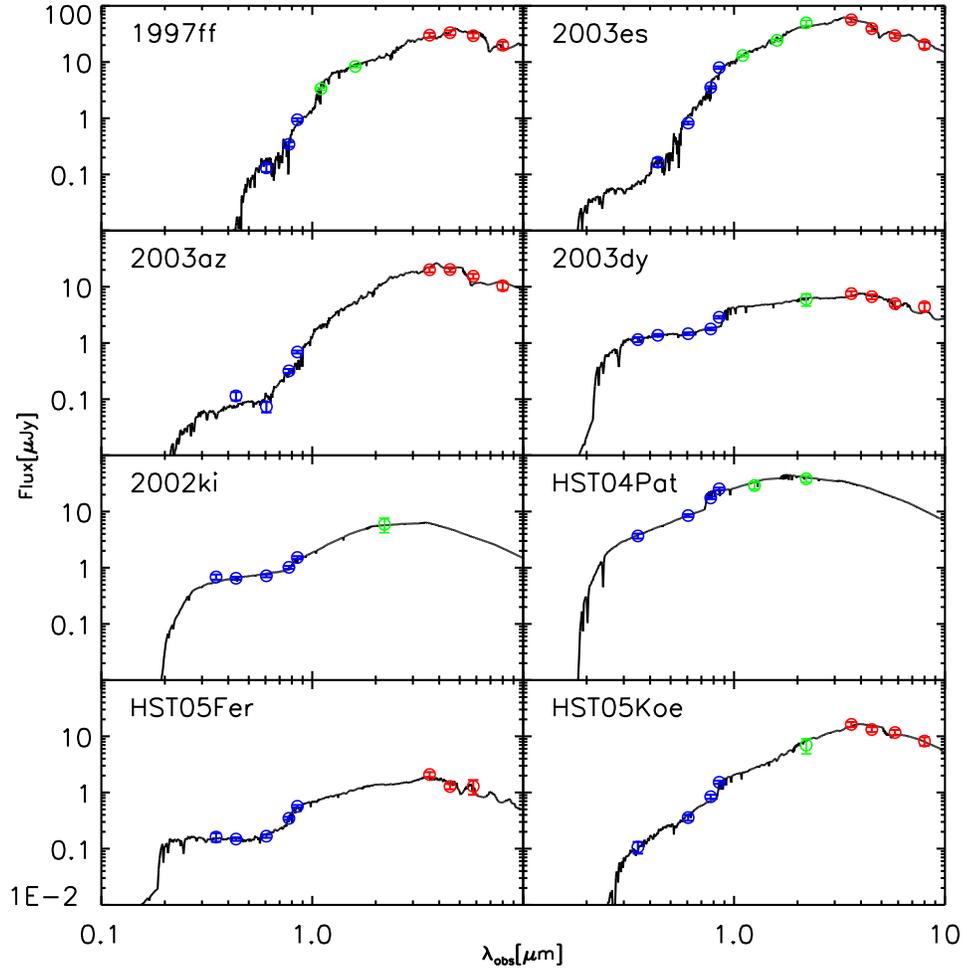}
 \caption[Best-fit SED plots]{SED plots for the best fit model for the SNe host galaxies, blue points are optical data, green points are near-infrared and red points are IRAC ch$1-4$.}\label{fig:seds}
 \end{figure}

\begin{figure}
   \centering
\includegraphics[width=12cm]{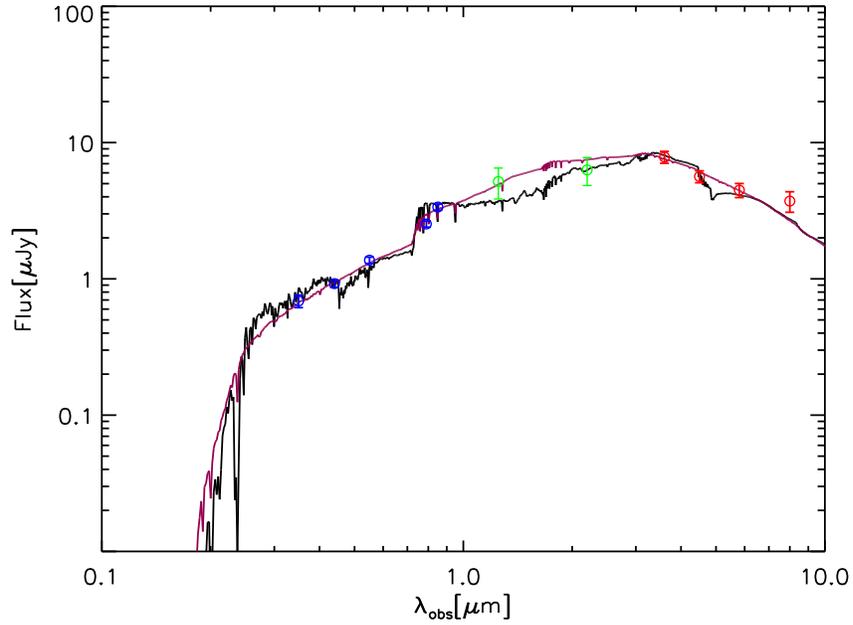}
 \caption[Effect of 24\micron Data]{SED plot for the best fit model for the host galaxy of SN\,Ia 1997fg, blue points are optical data, green points are near infrared and red points are IRAC ch$1-4$. The black line shows the best-fit model SED, the magenta line shows the best-fit model SED when the FIR luminosity limit
is not included. The best-fit luminosity-weighted age, A$_V$ combination changes from (0.02 Gyr, 2.1 mag) to (0.09 Gyr, 0.4 mag) when the limit is included.}\label{fig:no24}
 \end{figure}

We have found the best-fit stellar population model for each host
using a minimum $\chi^2$ technique. The set of best fitting parameters for each host along with the $\chi_\nu^2$ of the fit is shown in Table~\ref{tb:params}.  Since we have calculated the $\chi^2$ for each model, we calculate the errors by finding those bins with $\chi^2 < \chi^2_{min} + \chi^2_{gauss}$, where $\chi^2_{min}$ is the minimum $\chi^2$ of the fit and $\chi^2_{gauss}$ is the $\chi^2$ value for a 68\% and 95\% confidence level for a normal $\chi^2$ distribution with the same number of degrees of freedom as used in the fit (given by the number of photometric points less the number of parameters). The best-fitting SEDs along with the observed photometry used in the fit are shown in figure~\ref{fig:seds}.  It is also worth noting the effect of using the 24\,$\mu$m limit at this stage. Figure~\ref{fig:no24} is an enlarged version of the SED plot for 1997fg but also shown is the best-fitting SED found if the FIR luminosity limit is not used. The figure shows that both SEDs are reasonable
fits to the data (with $\chi^2$ of 11.1 when the limit is included and 3.6 when it is not), however by including the limit, the best-fit luminosity-weighted age changes from $\sim 0.02$ Gyrs to $\sim 0.09$ Gyrs. 

\begin{landscape}
\begin{deluxetable}{lllllllll}
\tabletypesize{\footnotesize}
\tablewidth{0pt}
\tablecolumns{8}
\tablehead{
\colhead{SN\,Ia} & \colhead{Age [Gyrs]} & \colhead{$\langle$Age$\rangle$ [Gyrs]} & \colhead{$A_V$[mags]} & \colhead{Z[Z$_\odot$]} & \colhead{$\tau$[Gyrs]} & \colhead{IMF} & \colhead{$\chi_{\nu}^2$} & \colhead{$M$ [log$_{10}$(M$_\odot$)]}}
\tablecaption{Table showing the best-fit parameters for each SN host (labeled by SN). Errors are calculated by finding the parameter value which raises the measured $\chi^2$(marginalised over all other parameters)  above the 68\% confidence threshold as calculated for figure~\ref{fig:stellarage}. In cases where the best-fit is at the extreme end of parameter space or if the marginalised $\chi^2$ never reaches larger than the threshold no error is given. $\langle$Age$\rangle$, Z and $M$ are the luminosity weighted age, metallicity and stellar mass respectively\label{tb:params}}
\startdata
1997ff & 0.571$^{+0.444}_{-0.062}$&0.561$^{+0.341}_{-0.090}$&0.5$^{+0.4}_{-0.3}$&2.500$_{-1.500}$ & 0.01$^{+0.09}$ & salpeter & 3.05&11.06$^{+0.150}_{-0.297}$\\
2003es & 2.300$^{+0.700}_{-1.161}$&1.999$^{+0.700}_{-0.898}$&0.3$^{+4.7}_{-0.3}$&1.000$^{+1.500}_{-0.995}$ & 0.30$_{-0.29}$ & chabrier & 1.95&10.80$^{+0.330}_{-0.328}$\\
2003az & 4.750$_{-3.316}$&3.897$^{+0.001}_{-2.763}$&0.7$^{+0.6}_{-0.1}$&2.500$_{-2.100}$ & 1.00$_{-0.70}$ & salpeter & 4.71&11.12$^{+0.301}_{-0.673}$\\
2003dy & 0.286$^{+4.214}_{-0.172}$&0.180$^{+1.568}_{-0.102}$&0.4$^{+0.2}_{-0.4}$&0.400$^{+2.100}_{-0.395}$ & 0.10$_{-0.09}$ & chabrier & 0.65&9.84$^{+0.731}_{-0.287}$\\
2002ki & 0.005$^{+4.995}_{-0.000}$&0.002$^{+1.961}_{-0.000}$&1.6$^{+3.4}_{-1.2}$&0.005$^{+2.495}$ & 0.01$$ & chabrier & 3.40&8.93$^{+1.291}_{-0.070}$\\
HST04Pat & 0.064$^{+0.017}_{-0.014}$&0.054$^{+0.017}_{-0.014}$&1.8$^{+0.2}_{-0.2}$&0.005$^{+2.495}$ & 0.01$$ & chabrier & 5.22&10.79$^{+0.288}_{-0.071}$\\
HST05Fer & 2.500$^{+3.000}_{-2.214}$&1.708$^{+0.484}_{-1.528}$&0.0$^{+0.7}$&0.400$^{+2.100}_{-0.200}$ & 1.00$_{-0.90}$ & chabrier & 1.83&9.23$^{+0.354}_{-0.381}$\\
HST05Koe & 0.072$^{+0.089}_{-0.008}$&0.062$^{+0.062}_{-0.008}$&1.8$^{+3.2}_{-0.2}$&2.500$_{-1.500}$ & 0.01$^{+0.02}$ & salpeter & 2.02&10.23$^{+0.145}_{-0.332}$\\
HST05Red & 0.055$^{+0.586}_{-0.023}$&0.031$^{+0.205}_{-0.009}$&0.8$^{+4.2}_{-0.4}$&1.000$^{+1.500}_{-0.995}$ & 0.03$_{-0.02}$ & chabrier & 2.55&8.84$^{+0.482}_{-0.111}$\\
HST05Lan & 1.700$^{+0.200}_{-0.091}$&1.400$^{+0.200}_{-0.091}$&0.3$^{+0.2}_{-0.3}$&1.000$^{+1.500}_{-0.600}$ & 0.30$_{-0.29}$ & salpeter & 1.66&10.66$^{+0.043}_{-0.298}$\\
HST04Tha & 2.200$^{+3.550}_{-1.061}$&1.899$^{+2.997}_{-0.798}$&0.0$^{+0.5}$&1.000$^{+1.500}_{-0.800}$ & 0.30$^{+0.70}_{-0.29}$ & chabrier & 2.02&10.07$^{+0.548}_{-0.275}$\\
HST04Eag & 5.500$_{-5.436}$&2.192$^{+3.298}_{-2.138}$&0.6$^{+0.2}_{-0.3}$&0.005$^{+2.495}$& Constant SFR & salpeter & 2.49&10.48$^{+0.476}_{-0.997}$\\
HST05Str & 0.072$^{+0.072}_{-0.008}$&0.062$^{+0.045}_{-0.008}$&0.6$^{+4.4}_{-0.1}$&2.500$_{-1.500}$ & 0.01$^{+0.02}$ & chabrier & 1.53&9.25$^{+0.353}_{-0.064}$\\
1997fg & 0.102$^{+2.898}_{-0.102}$&0.092$^{+2.088}_{-0.092}$&0.4$^{+0.1}_{-0.3}$&2.500$_{-2.495}$ & 0.01$^{+0.99}$ & chabrier & 1.86&9.43$^{+0.860}_{-0.086}$\\
2003aj & 0.161$^{+4.339}_{-0.108}$&0.124$^{+1.624}_{-0.082}$&0.1$^{+0.9}_{-0.1}$&1.000$^{+1.500}_{-0.995}$ & 0.03$_{-0.02}$ & salpeter & 0.77&9.25$^{+0.611}_{-0.335}$\\
2002fx & 2.750$^{+1.500}_{-2.746}$&1.225$^{+0.419}_{-1.224}$&0.0$^{+1.2}$&0.005$^{+2.495}$& Constant SFR & chabrier & 1.63&8.94$^{+0.373}_{-0.875}$\\
2003ak & 0.128$^{+0.127}_{-0.078}$&0.118$^{+0.035}_{-0.078}$&0.0$^{+0.6}$&1.000$^{+1.500}_{-0.995}$ & 0.01$^{+0.09}$ & chabrier & 0.53&9.70$^{+0.435}_{-0.160}$\\
2002hp & 1.700$^{+2.800}_{-0.795}$&1.587$^{+2.903}_{-0.720}$&0.3$^{+0.5}_{-0.3}$&1.000$^{+1.500}_{-0.800}$ & 0.10$^{+0.20}_{-0.09}$ & salpeter & 0.45&11.00$^{+0.379}_{-0.459}$\\
HST04Mcg & 0.128$^{+3.122}_{-0.047}$&0.118$^{+2.300}_{-0.047}$&1.5$^{+0.1}_{-1.5}$&1.000$^{+1.500}_{-0.995}$ & 0.01$^{+0.99}$ & salpeter & 0.56&10.61$^{+0.508}_{-0.372}$\\
HST04Omb & 0.203$^{+0.516}_{-0.153}$&0.108$^{+0.158}_{-0.068}$&0.0$^{+0.2}$&1.000$^{+1.500}_{-0.600}$ & 0.10$_{-0.09}$ & chabrier & 1.27&9.14$^{+0.345}_{-0.157}$\\
\enddata
\end{deluxetable}
\end{landscape}


  \begin{figure}[htb]
  \centering
\includegraphics[width=14cm]{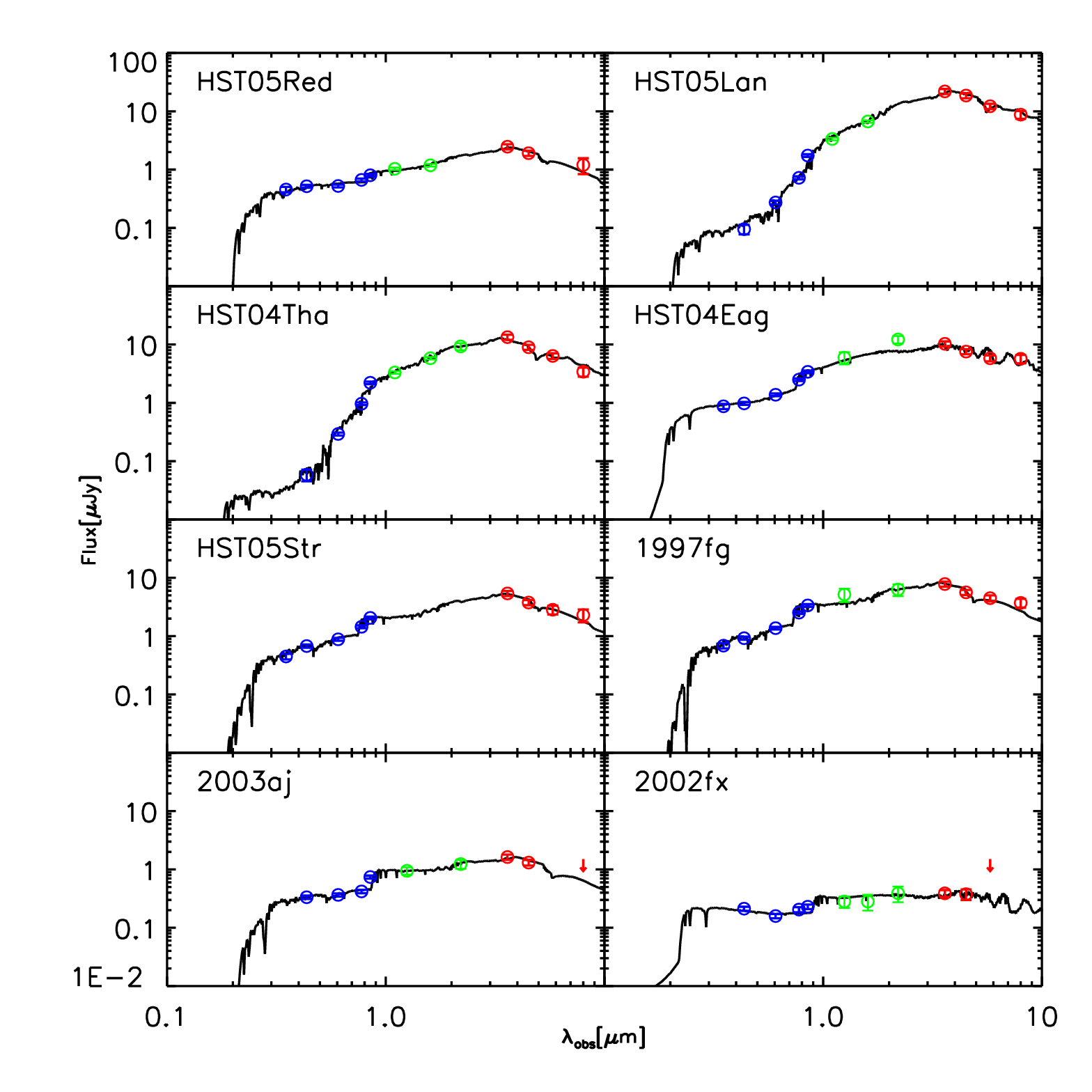}
\begin{center}
\emph{figure~\ref{fig:seds} continued}
\end{center}
 \end{figure}

  \begin{figure}[htb]
  \centering
\includegraphics[width=14cm]{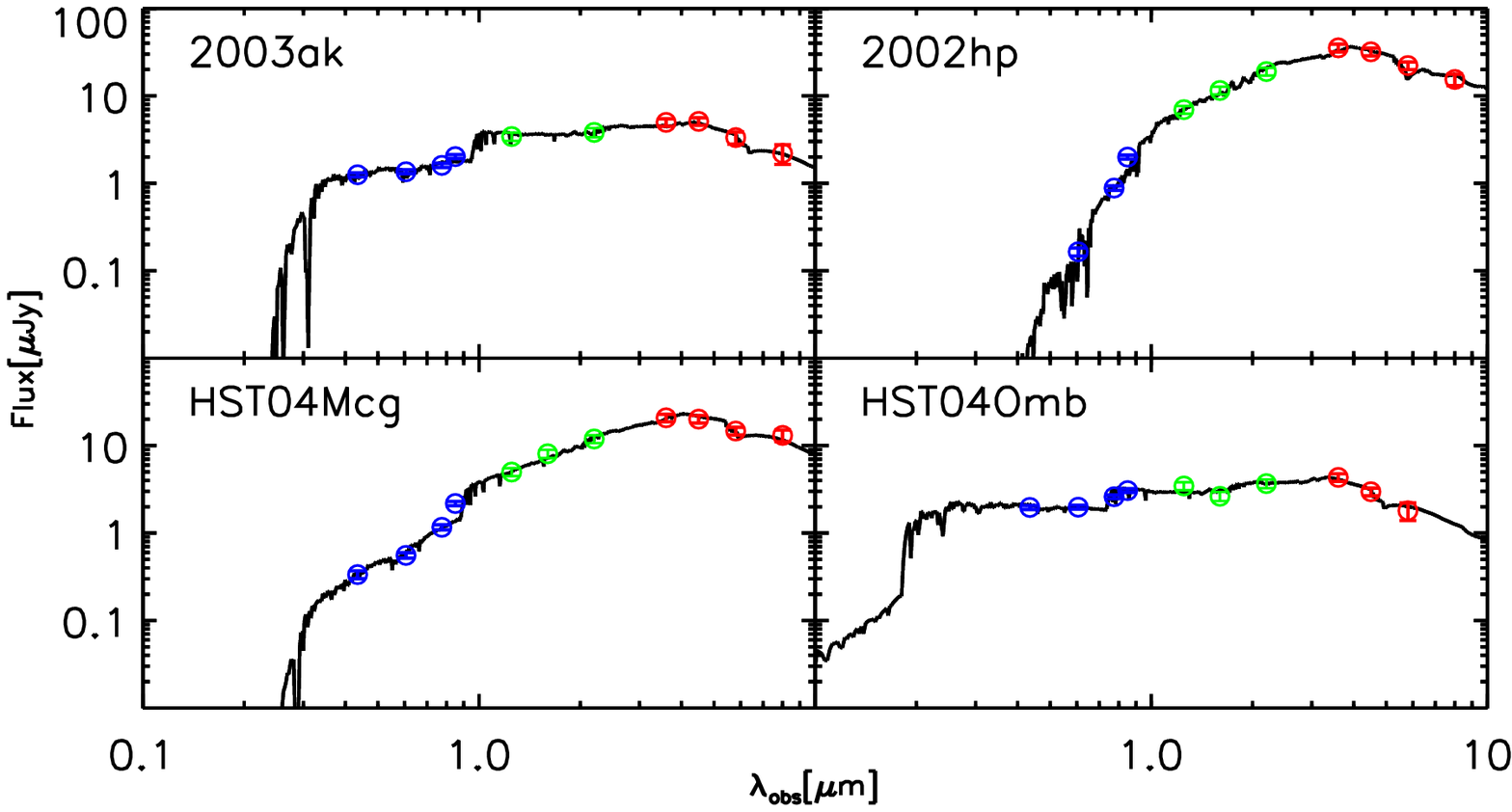}
\begin{center}
\emph{figure~\ref{fig:seds} continued}
\end{center}
 \end{figure}
 
 \clearpage

Since we are only fitting one population to the photometry of the host galaxies it is possible that the SN\,Ia progenitors were
born in a more recent starburst which only contributes a fraction to the total emission of the galaxy which is dominated by an older population. In this case we would overestimate the ages as we would fit to the dominant, older population but the SN\,Ia in fact comes from a younger population. In order to test this possibility we perform a two-component fit. We use the best-fitting parameters for metallicity, SFH and IMF and take a model which is as old as the universe at the host redshift and a model which is 10 Myrs old. We then add the two models together allowing the fractional contribution from the young population to vary between 0 - 1 in steps of 0.1. We also allow the extinction to vary and to be different in the two populations.  In all cases the resulting fit is worse than the original best-fit using only one population. This suggests that our assumption that the SNe\,Ia most likely come from the one population model we are fitting is valid. 

We also assessed the \citet{charlot00} prescription for the dust extinction. We find that the ages using the \citet{charlot00} recipe differ from those using the \citet{calzetti00} recipe  by more than a factor of 2, in only 3 cases.  In two of these cases these ages are larger than those using the \citet{calzetti00} recipe, however, in all 3 cases the $\chi^2$ of the best-fit is lower using a \citet{calzetti00} recipe compared to the \citet{charlot00} recipe.  We conclude that our derived ages are robust with respect to the fitting technique.

\begin{figure}[htb]
\centering
\includegraphics[width=12cm]{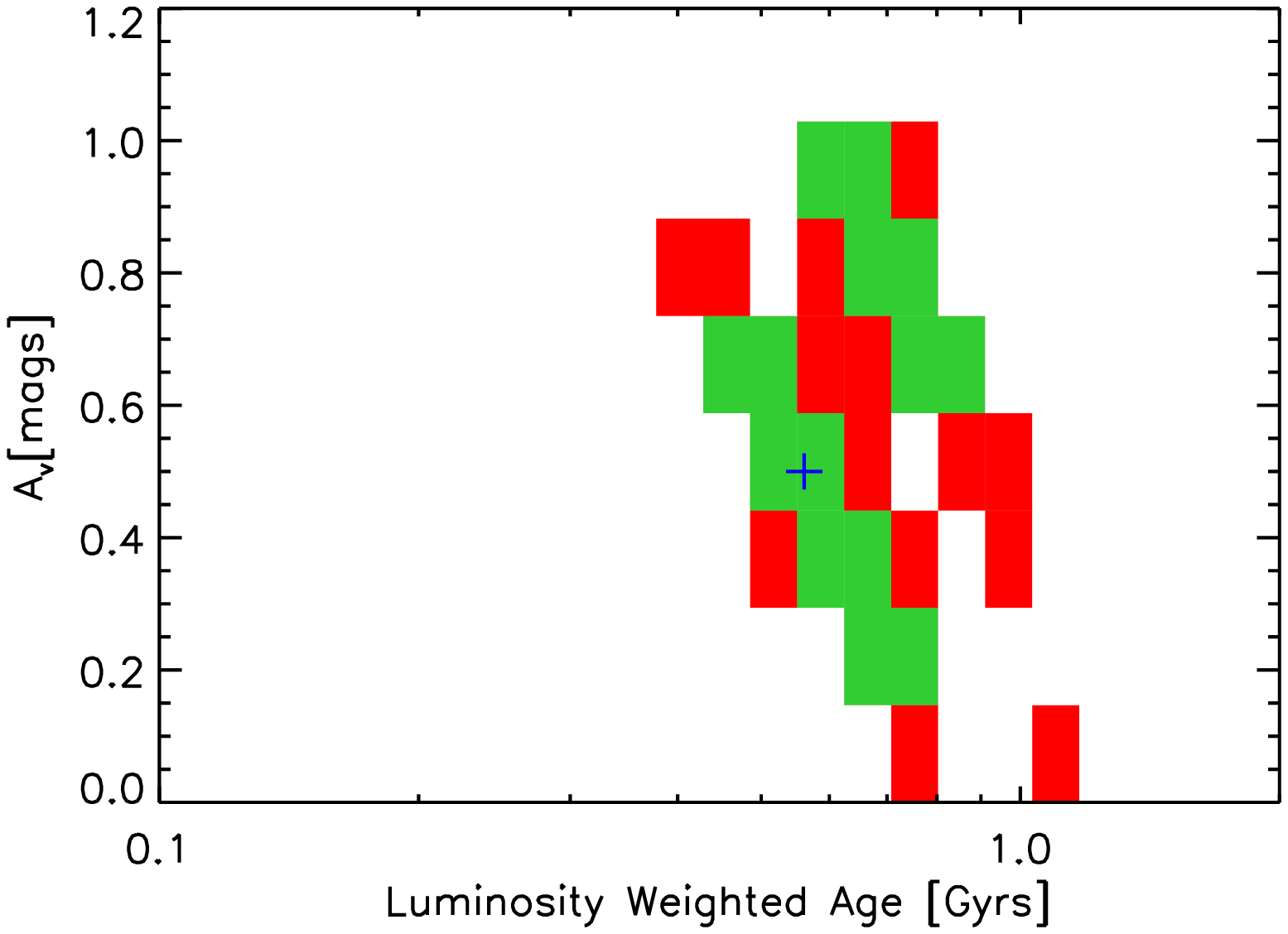}
\includegraphics[width=12cm]{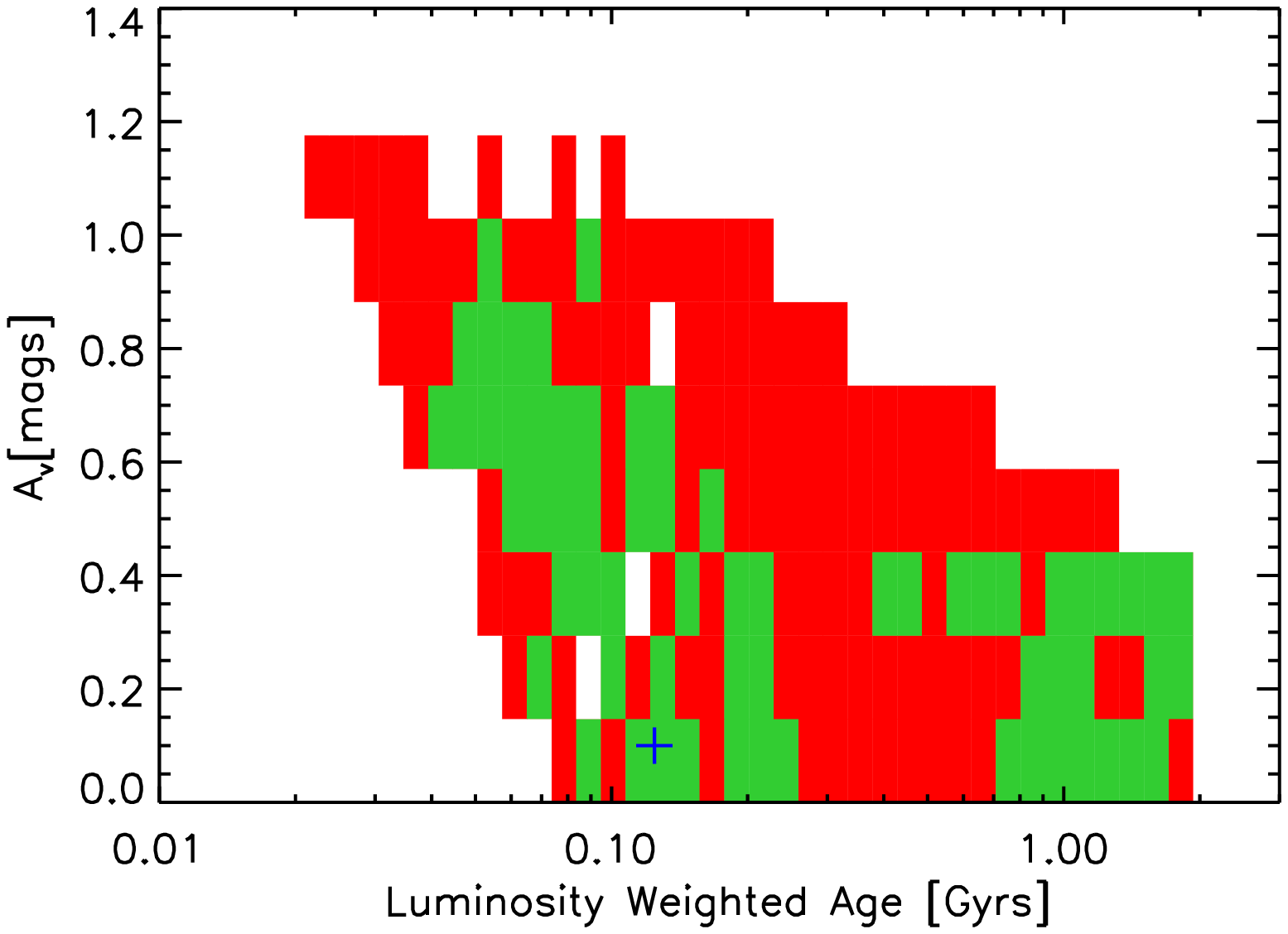}
\caption[SED Fitting Age-$A_V$ Confidence Levels]{Contour plots of the $\chi^2$ distribution for two representative SNe\,Ia host galaxies (top: 1997ff, bottom: 2003aj).  The contours were calculated by collapsing over the remaining axes, i.e. for a fixed pair of age-$A_V$ values the minimum $\chi^2$ allowing all other parameters to vary was found.  This was then binned together. The green area is the 68\% confidence level and the red area is the 95\% confidence level. The confidence levels were found by finding those bins with $\chi^2 < \chi_{min}^2 + \chi_{gauss}^2$, where  $\chi_{min}^2 $ is the minimum $\chi^2$ across the whole parameter range (i.e. the $\chi^2$ of the best-fit) and $ \chi_{gauss}^2$ is the $\chi^2$ value which gives the 68\% or 95\% confidence level for a normal $\chi^2$ distribution with the same number of degrees of freedom as used in the fit. The blue cross gives the position of the best fit. 
}\label{fig:contours}
 \end{figure}

Figure~\ref{fig:contours} shows
contour plots of the $\chi^2$ distribution in the age - $A_V$ plane for the SNe host galaxies, showing 68\% and 95\% confidence intervals. The figure shows that in some cases there is a fairly smooth distribution of the degeneracy between the age of the stellar population and the dust extinction, in that good fits can be achieved with younger populations with a higher dust extinction, as expected. This degeneracy remains despite our removing any models which give a FIR luminosity that is inconsistent with the MIPS 24$\mu$m data, although it is much reduced. Many panels of figure~\ref{fig:contours}, however, do not show a  smooth distribution. While they generally show the same degeneracy, it is clear that the age parameter is not sampled sufficiently, in particular at  older ages to give a smooth distribution. Unfortunately, we are unable to alter the ages of the output of the stellar population models. However, the uncertainties in table~\ref{tb:params} and figures~\ref{fig:stellarage}, ~\ref{fig:epochsf} 
span the entire range of ages that give consistent fits.

\begin{figure}[htb]
 \centering
 \includegraphics[width=12cm]{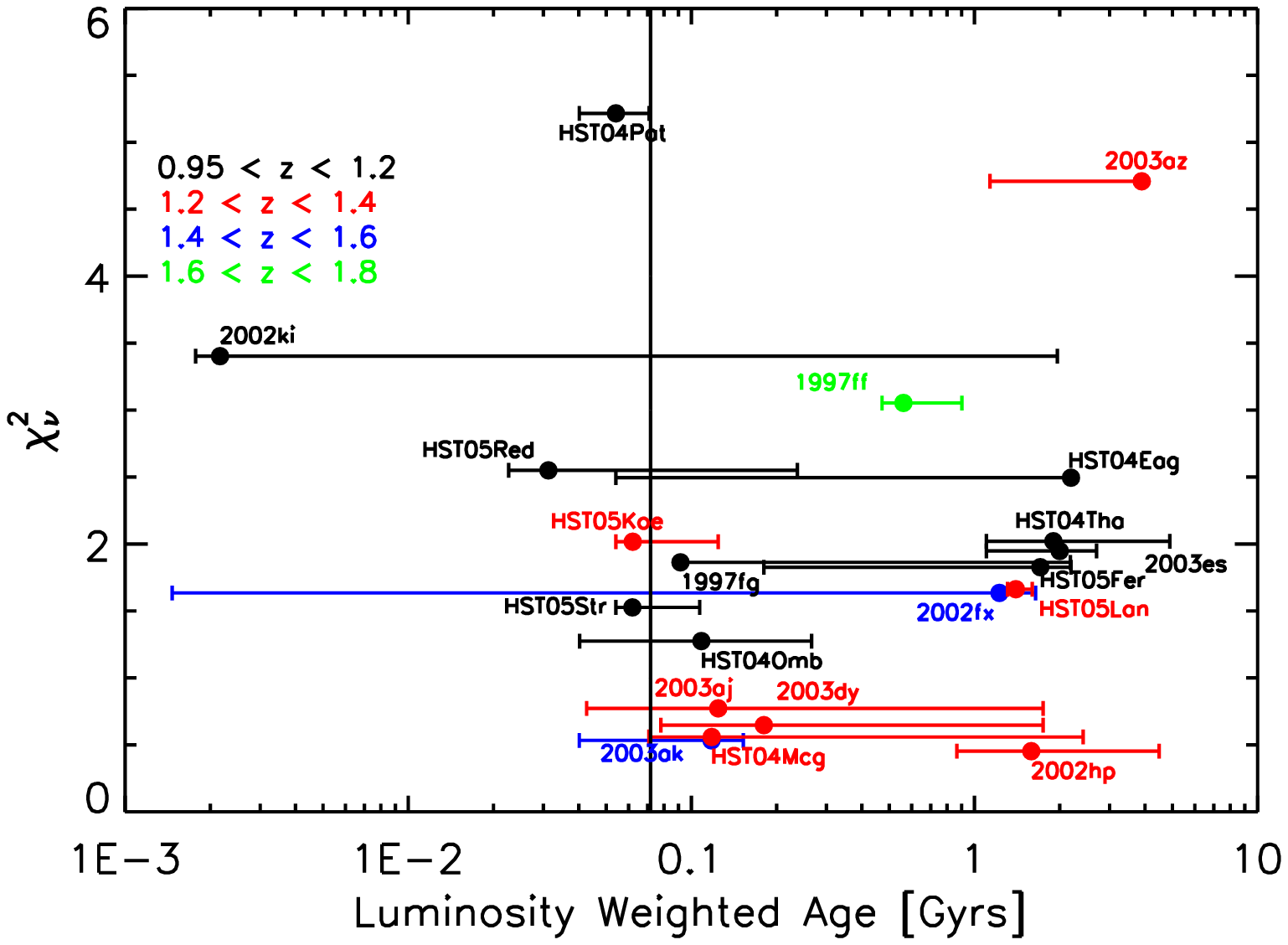}
 \caption[Luminosity-Weighted Stellar Population Ages]{The plot shows the luminosity-weighted ages of the stellar populations in the hosts with the associated  $\chi_{\nu}^2$. There are some SNe\,Ia which originate from very young hosts ($<0.1$Gyr). The vertical line is the weighted mean age of 0.07 Gyrs.
 }\label{fig:lumstellarage}
 \end{figure}

\begin{figure}[htb]
 \centering
 \includegraphics[width=12cm]{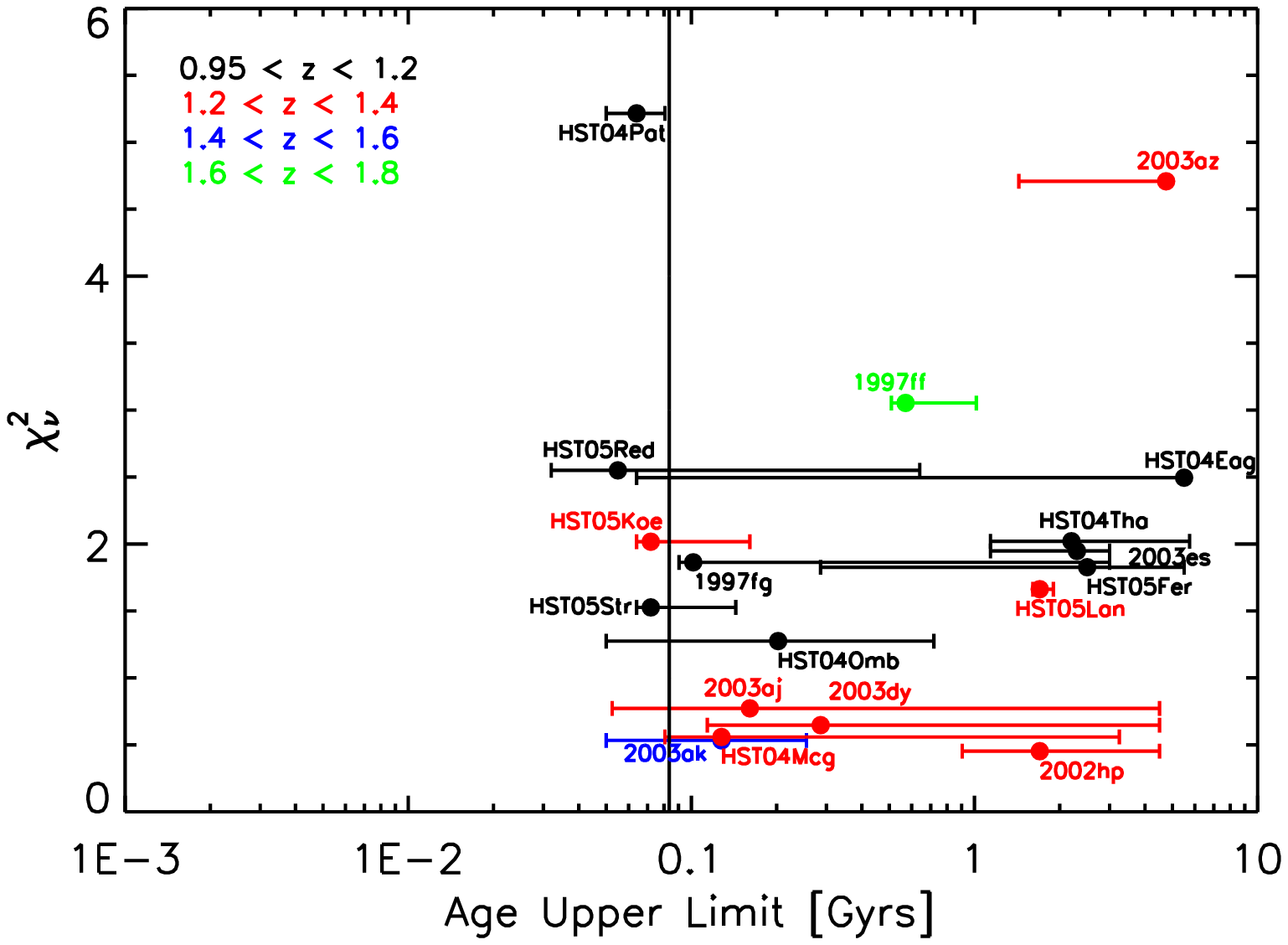}
 \caption[Upper-Limit Stellar Population Ages]{The plot shows figure~\ref{fig:lumstellarage} but when using the age upper-limits rather than luminosity-weighted ages. The vertical line is the weighted mean age of 0.08 Gyrs.}
 \label{fig:stellarage}
 \end{figure}

 Figure~\ref{fig:lumstellarage} shows the luminosity weighted stellar population ages of the host galaxies. This age is plotted for each SN\,Ia host with the reduced $\chi^2$ ($\chi_\nu^2 = \chi^2/\nu$, where $\nu$ is the number of degrees of freedom) of the best-fit model. The error bars are calculated from the grid of $\chi^2$, whereby we find the age at which the $\chi^2$ rises above the 68\% confidence level, collapsing over all other parameters. The plot shows that there are two hosts with exceptionally large error bars, namely the host galaxies of  2002fx and 2002ki. In the latter case this is most likely due to the lack of IRAC photometry, in the former case this is most likely due to large photometric uncertainties. In any case we remove these hosts from the analysis (including the calculations based on this plot). The plot shows a large range of luminosity weighted ages from 0.03 - 3.90 Gyrs, with both prompt and delayed SNe\,Ia. As we wish to constrain the maximum delay time, we show the plot again but when the upper-age limit (i.e. the best fitting time-step of the CB07 models) rather than the luminosity-weighted mean age is used in figure~\ref{fig:stellarage}. Both plots show that the weighted mean age of the SNe hosts is $<0.1$ Gyr implying that they harbor young stellar populations. This fact was noted earlier 
by \citet{chary05}
who found that $\sim$60\% of the SN Ia host galaxies harbor young, dusty starbursts based on their detection at 24\,$\mu$m; this is a factor
of 1.5 higher than the field galaxy population.

\begin{figure}[htb]
  \centering
 \includegraphics[width=12cm]{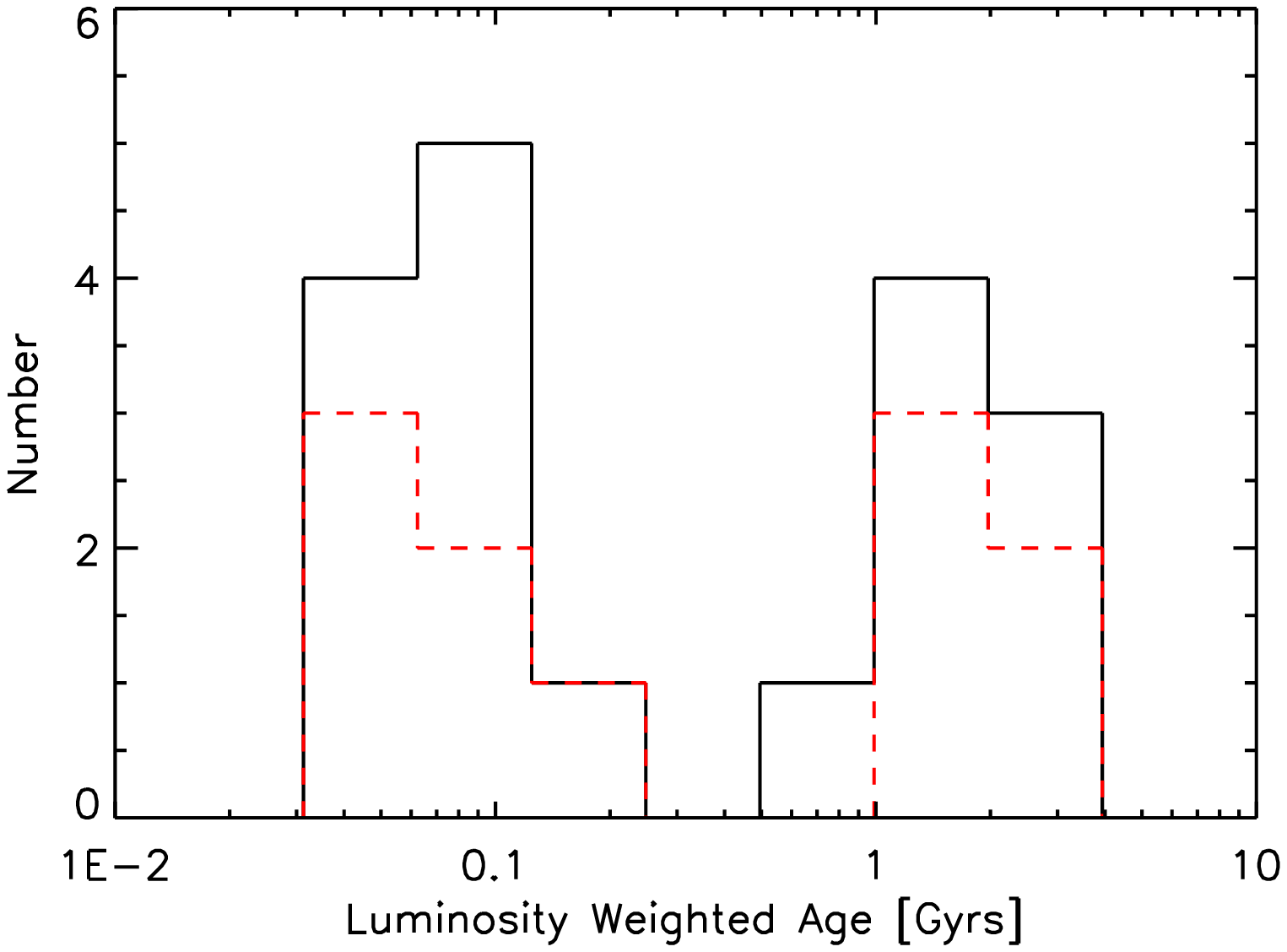}
 \caption[Histogram of Luminosity Weighted Ages]{A histogram of the luminosity weighted ages of the stellar population in $z \ge 0.95$ Type Ia SN host galaxies. The ages appear to show a bi-modal distribution with the younger population having a median age of 0.11 Gyrs and the older population having a median age of 1.9 Gyrs. The black solid line shows the distribution for all SNe and the red dashed line shows the distribution when only high confidence SNe\,Ia are included. 
 }\label{fig:lumagedist}
 \end{figure}

The distribution of the best-fit luminosity-weighted stellar population ages are shown in figure~\ref{fig:lumagedist}. We show the histogram when only high confidence SNe\,Ia are included, these are SNe which are classed as `Gold' in \citet{riess07} and also have good spectra of the SNe (red-dashed line). The black solid line represents all SNe and includes `Silver' SNe and `Gold' SNe without spectra. `Bronze' SNe are not included in any of this analysis. The figure suggests a bi-modal distribution with median ages of 0.11 and 1.9 Gyrs for the two populations (both the whole sample and when only considering the high-confidence SNe\,Ia). We arbitrarily consider a host age $<0.4$ Gyrs to be young, as this is the minimum in the distribution and since there is no precise definition in the literature. We perform a KS-test to determine whether a bi-modal distribution is a good description of the data. 
Comparing to a distribution comprised of two gaussian distributions centered at the young and old median ages (allowing the standard deviation of each gaussian  and the ratio between their amplitudes to vary) gives a KS-statistic of 0.99 for the high confidence \sneia  and 0.97 for the whole sample.
The KS statistic for the distribution to be fit by a single Gaussian has a probability of only 42\%. Thus, a bimodal distribution is clearly favored by
the age distribution although a larger sample of objects such as those available from the upcoming CANDELS/WFC3 survey is 
required for definitive confirmation of a bimodal distribution of ages. 
 
Due to the large error bars for the ages of the stellar populations we perform a monte carlo simulation to test the strength of any bi-modality. 
For each simulation, we calculate an age for each SN from the probability distribution calculated from the $\chi^2$ distribution of the error bars shown in figure~\ref{fig:lumstellarage}. We then repeat the KS-test. We perform this simulation 1000 times. We find that a bi-modal distribution is preferred 95\% and 97\% of the time for the whole sample and for the high confidence sample respectively. Finally, we use the simulation to test the existence of prompt SNe given the large error bars of figure~\ref{fig:lumstellarage}. The mean number of young hosts in the simulation is $9.5\pm0.05$ and of old hosts is $10.5\pm0.05$ for the full sample. When considering only the high confidence SNe\,Ia the average number of young and old hosts is  $4.6\pm0.03$ and $6.4\pm0.03$. These results suggest that we can be fairly confident of the existence of prompt Type Ia SNe at $z\gtrsim1$. 

\begin{figure}[htb]
\centering
\includegraphics[width=12cm]{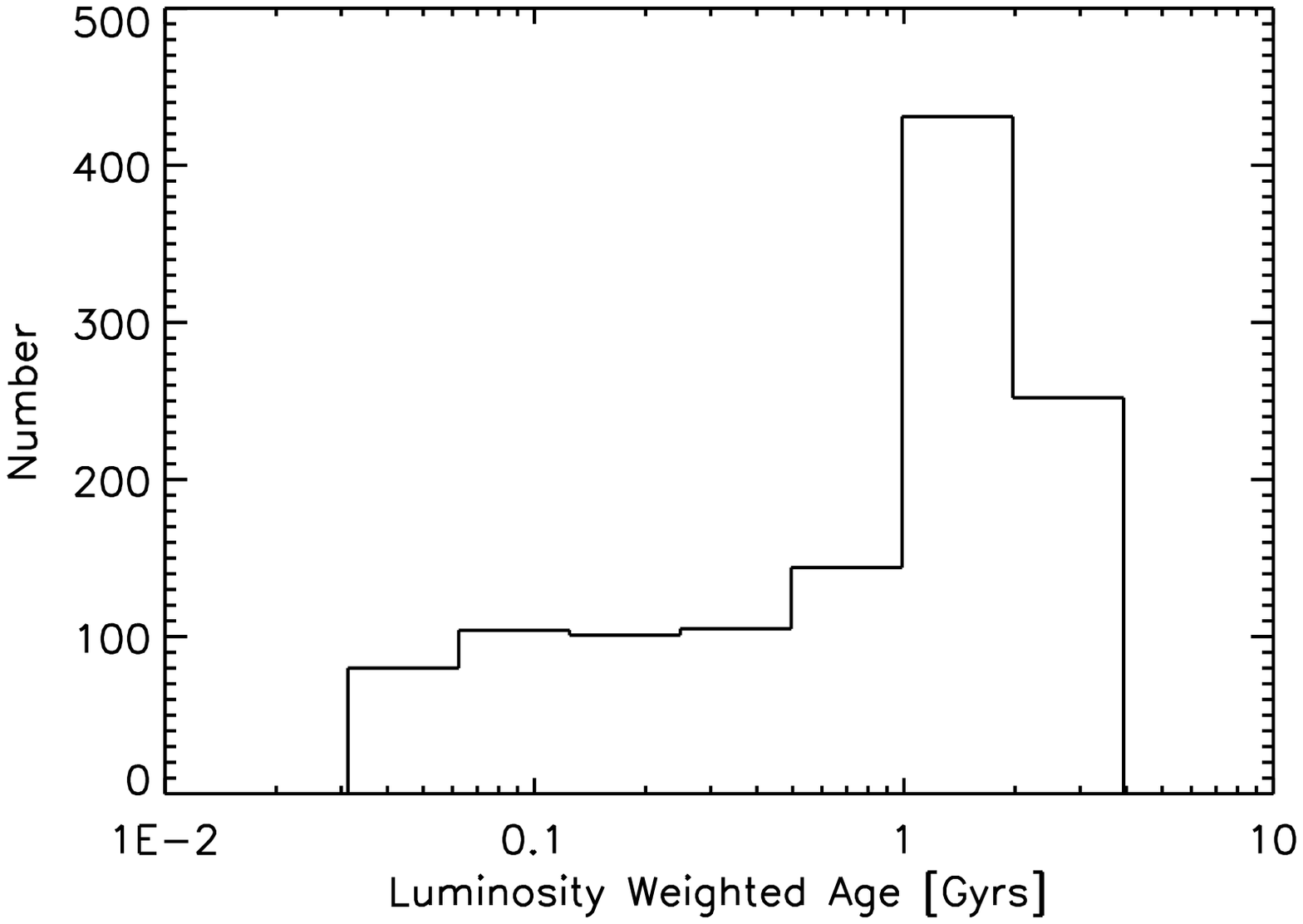}
\caption[Field Sample Luminosity Weighted Ages]{Histogram of the luminosity weighted ages of the field galaxy sample in the same redshift range as the SNe hosts, used in the DTD calculation.}\label{fig:fieldages}
\end{figure}

\begin{figure}[htb]
\centering
\includegraphics[width=12cm]{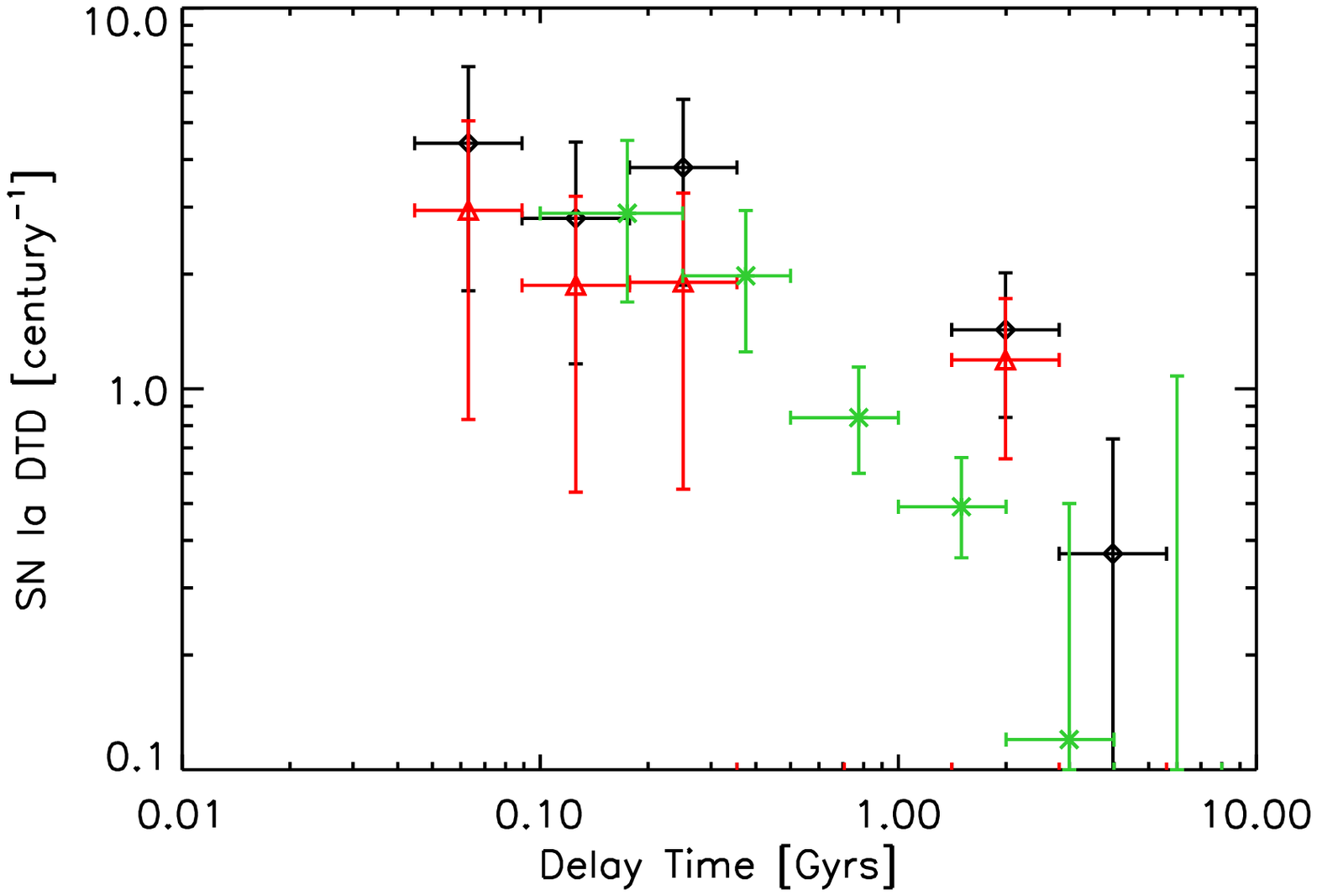}
\caption[SNe\,Ia Delay Time Distribution]{Type Ia Supernova Delay Time  Distribution (DTD) for $z\gtrsim1$ SNe. The black points are those obtained for the whole sample, the red points are that obtained for the high confidence SNe only. Also shown is the DTD obtained by \citet{totani08} in green. 
Although, the uncertainties are dominated by the small number of candidates, if
the DTD were a smooth power-law, we would have expected to find at least two additional
SNe hosts with delay times in the 0.2-0.6 GYr range.}\label{fig:dtd}
\end{figure}

Finally, we wish to calculate the delay time distribution (DTD). In order to do this we must account for the selection efficiency of the supernova search. Typically, this is achieved in the form of a `control time' calculation (which is the total time that a SN could have been detected). We use the control time calculation for this SNe search of \citet{dahlen08} as an estimate of the probability of detecting a supernova as a function of redshift. \citet{dahlen08} give supernova rates and the number of supernovae in four redshift bins between 0.2 and 1.8. We use the three highest redshift bins to interpolate the control time at each of the redshifts of our SNe using the equation
\[
R = \frac{N}{t_c \Delta V}
\]
\noindent where $R$ is the SNe\,Ia rate, $N$ is the number of SNe observed in a redshift bin, $t_c$ is the control time and $\Delta V$ is the volume of the redshift bin \citep{strolger10}. We then divide the best-fit luminosity-weighted stellar population ages by this control time. In order to calculate the DTD we then calculate a histogram of this distribution and divide the histogram by the equivalent distribution of a sample of the field population in the GOODS survey with spectroscopic redshifts and $z\ge0.95$. The field sample we use consists of 1507 galaxies across the GOODS-N and GOODS-S fields. Figure~\ref{fig:fieldages} shows a histogram of these ages and figure~\ref{fig:dtd} shows the resulting DTD. We again perform a monte carlo simulation of the DTD. We find that a bi-modal distribution is preferred 80\%  of the time for the whole sample and 74\% of the time when considering only the high confidence \sneia. For the whole sample an exponential distribution is preferred 11\% of the time and a single gaussian distribution 8.7\%. For the high-confidence sample a single gaussian is prefered in 14\% of the simulations and an exponential distribution 7\% of the time. In figure~\ref{fig:dtd} we also plot the DTD obtained by \citet{totani08}. For most bins the two measurements are consistent, however, between 0.4 and 2 Gyrs  the \citet{totani08} result has a larger SNe rate than that found here and their results do not appear to be bi-modal. The \citet{totani08} result was obtained over a lower redshift window, extending between $0.4<z<1.2$ and it is possible the difference reflects a change in the dominant SNe\,Ia progenitor.

\subsection{Evidence Against Truncation of the High Redshift Stellar IMF}
\begin{figure}[htb]
 \centering
 \includegraphics[width=12cm]{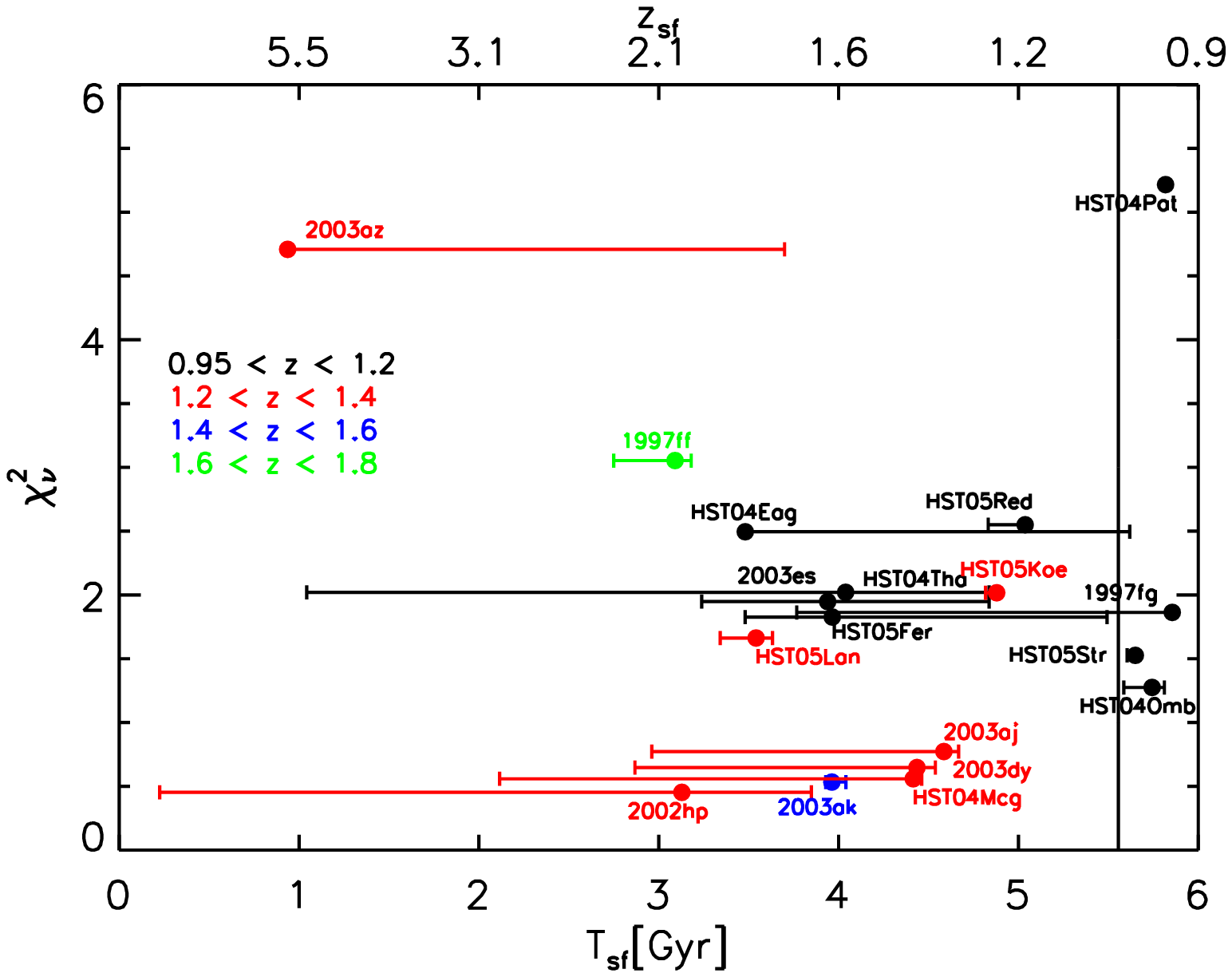}
 \caption[First Epoch of Low-Mass Star Formation]{The plot shows the first epoch of low mass star formation for each host with the associated $\chi_{\nu}^2$. The vertical line is the weighted mean epoch of low mass star formation of 5.6 Gyrs. The $\lesssim8 $M$_\odot$ progenitor stars of SNe\,Ia are certainly in place by $z \sim 2$ and possibly by $z \sim 5$. 
 }\label{fig:epochsf}
 \end{figure}

Since SNe\,Ia are believed to be the explosions of White Dwarfs (WD) which have reached the Chandrasekhar mass, the stars eventually giving rise to these explosions are low mass stars. A star of $\gtrsim$8\,$M_{\odot}$ will explode as a core-collapse supernova instead of forming a WD \citep{blanc08}. Therefore, with the ages of the underlying stellar populations which produce the SNe\,Ia we can constrain the first epoch of low mass star formation, $T_{sf}$
\[
T_{sf} = T_{z} - T_{stellar}
\]
where $T_{z}$ is the age of the universe at the
redshift of the host galaxy and $T_{stellar}$ is the best-fit luminosity-weighted stellar population age. $T_{sf}$ is the time since the Big Bang when the stellar population formed. The results are shown in figure~\ref{fig:epochsf}  where the error bars are calculated as above. The figure suggests that $\lesssim 8 \textrm{M}_{\odot}$ stars formed within 3 Gyrs of the Big Bang and possibly by $z \sim 5$. If these results are confirmed they are in contrast to the proposal by \citet{tumlinson04} who suggest that instead of requiring the first stars to be very massive stars ($M > 140\textrm{M}_\odot$) the primordial IMF
may be truncated at $\sim 10-20$M$_\odot$ at $z \gtrsim 6$. This suggestion was primarily
based on the Fe-peak and $r-$process elemental abundance patterns of extremely metal poor stars in the Galactic halo.

Interestingly, we do not see any trend of peak absolute magnitude with host age; based on the previous work we would expect younger galaxies to host fainter \sneia. However, as pointed out by \citet{sullivan10} the sample used here is perhaps too small to see the slight trend found by those authors. Furthermore, \citet{neill09}, found that for a sample of local SNe\,Ia that there was no trend with light-curve corrected peak $B$-band absolute magnitude and host luminosity weighted age except when a subset of hosts with low dust extinction was considered.


\section{Discussion}

We have utilised the multi-wavelength spectral energy distribution of the host galaxies of Type Ia SNe at $z\gtrsim1$ to identify a possible bi-modal distribution
of luminosity-weighted stellar ages and thereby delay times between the burst of star-formation and the time at which the SN explodes. We find evidence for both prompt (i.e. short delay times, $\lesssim0.4$ Gyrs) and delayed (i.e. long delay times) SNe\,Ia, with some extremely young ($\lesssim0.1$Gyrs) luminosity weighted ages.  We discuss our results in the context of past measurements at low and high redshift.

\subsection{Low Redshift}

In the low redshift universe there is evidence for both prompt and delayed SNe\,Ia. \citet{aubourg08} find significant evidence for a population of SNe\,Ia with progenitor lifetimes of $<$0.18 Gyrs. \citet{mannucci05} find that SNe\,Ia are more common in blue rather than red galaxies and \citet{dellavalle05} find that the SN\,Ia rate is 4 times higher in radio-loud rather than radio-quiet galaxies, suggesting 
that SNe\,Ia are associated with younger stellar populations and therefore shorter delay times. \citet{schawinski09} fit SDSS and GALEX photometry of the host galaxies of 21 local SNe\,Ia in early type galaxies to a two-component  Stellar Population  model, with an old component (of age varying 1 - 15 Gyrs) to represent the older, underlying population and a young starburst component with varying age and mass fraction. They find no \sneia with delay times $<0.1$ Gyrs, and a range of minimum delay times of 0.275 - 1.25 Gyrs. This is perhaps to be expected as only SNe\,Ia in early type galaxies are studied. Furthermore, measurements of the SN\,Ia rate at different redshifts have suggested that the delayed SNe\,Ia give a more significant contribution to the total rate at low redshift \citep{sullivan06,neill06,neill07}.

\citet{gallagher05} perform a spectroscopic study of the host galaxies of 57 local SNe\,Ia to deduce the SFR and SFH of the host galaxy. By measuring the Scalo-b parameter they see evidence for a bi-modal distribution which further suggests two progenitor classes for the SNe. However they put a lower limit on the delay time of 2 Gyrs. It is hard to see how to reconcile these two results, although again these are local galaxies when we would expect more delayed SNe than prompt SNe.

\citet{neill09} perform an SED fitting analysis of UV and optical photometry of the host galaxies of a sample of local SNe\,Ia. They confirm the results of \citet{sullivan06} who showed that brighter SNe occur in galaxies with higher specific SFR.  \citet{gallagher08} obtained optical spectra for the host galaxies of 29 SNe\,Ia selected to be local early type galaxies.  From comparisons to stellar population synthesis models they find a correlation between age or metallicity with peak SN\,Ia $V-$band absolute magnitude, preferring a trend with age (based on the trend of SN\,Ia rate with specific SFR) such that SNe\,Ia from older populations are fainter. \citet{howell09} find a very weak correlation between luminosity-weighted age of the host and $^{56}$Ni mass derived from the integrated luminosity of the SN also suggesting that older, low mass progenitors produce fainter SNe\,Ia. 

\subsection{High Redshift}
Many authors have attempted to constrain the delay time distribution (DTD) of SNe\,Ia by comparing the observed SN\,Ia rate to that predicted by a convolution of the DTD with an assumed SFH, \citep{galyam04,strolger04,dahlen04,barris06,dahlen08,strolger10}. These studies have argued for a range of
characteristic delay times, spanning 1 - 4 Gyrs.

This seems at odds with our results which show a large proportion of SNe\,Ia with delay times $<0.1$ Gyrs. Indeed, the \citet{strolger04,strolger10} result, finding a characteristic delay time of $3-4$ Gyrs, is based on an analysis of the same set of SNe used in the present work.  \citet{dahlen04,dahlen08} use the models of \citet{strolger04} with a measurement of the GOODS SN\,Ia rate to show that the best fit DTD is a  Gaussian with a  mean delay time of 3.4 Gyrs. However, \citet{forster06,blanc08} have shown that the results of such analyses are strongly dependent on the assumed SFH, introducing systematic errors, and as such our results are not necessarily at odds with those authors. Results from \citet{oda08} which attempt to fit both the SFH and DTD simultaneously are only able to put weak constraints on the DTD. Furthermore, \citet{poznanski07} measure the SN\,Ia rate at a similar redshift range using a dataset from the Subaru Deep Field and find a more constant rate which could suggest shorter delay times. In a companion paper to \citet{galyam04}, \citet{maoz04} show that iron abundances in clusters require delay times of $<2$ Gyrs. 

These seemingly contradictory results, with evidence for both prompt and delayed SNe\,Ia at high and low redshift have led to the suggestion of a two-component DTD \citep{mannucci05}. For example, \citet{mannucci06} use several datasets to show that the observations cannot be simultaneously matched by a single delay-time. Several authors have developed this further suggesting a  model  with a prompt component dependent on the specific SFR (SFR per unit stellar mass) of the host galaxy and a delayed component dependent on the stellar mass of the host galaxy \citep{scannapieco05,sullivan06,neill06,neill07,aubourg08}.  These DTDs are then matched to the observed SN\,Ia rate to fit the model parameters, tending to find a best-fit DTD dominated by the prompt component, especially at high-redshift. Furthermore, in many of these models the contribution of the prompt component is expected to increase with redshift \citep{sullivan06}. The average SNe\,Ia light curve width appears to increase with redshift, supporting these models \citep[prompt SNe\,Ia are more luminous and have a broader light curve][]{howell07,sullivan09}.
However, some studies have shown that SN\,Ia rate measurements are unable to differentiate between DTD models to any significance \citep{neill06,neill07,blanc08,botticella08,oda08} and that these results are still highly dependent on the choice of SFH. \citet{kuznetsova08} re-computed the results of \citet{dahlen04} using a more sophisticated technique and additional data to find that they could not discriminate between a two-component model and a gaussian single delay time DTD.

\citet{totani08} perform an analysis similar to that presented here using SN\,Ia host galaxies in the Subaru/XMM-Newton Deep Survey (SXDS) in the redshift range $0.4 - 1.2$. They find a DTD in the range $0.1 - 8$ Gyrs (extending to 10 Gyrs from the SN\,Ia rate in local ellipticals). This analysis selects old, passively evolving galaxies and therefore does not probe the shortest delay times. However, they do not find a bi-modal distribution as we find here.

There are a number of implications of this result for cosmological studies. That there is a large population of \sneia which have short delay times means that these objects could be used as cosmological probes to very high redshift ($z\gtrsim3$). Furthermore, if prompt and delayed \sneia have  different light-curve shape/luminosity relations the host galaxy will need to be taken into account when using \sneia to determine the equation of state parameter of dark energy, $w$, especially when considering possible evolution of this parameter. Using a large dataset of \sneia spanning a wide redshift range \citet{sullivan10} showed that \sneia light-curve widths depend on host galaxy specific star formation rate (SSFR) and stellar mass, with narrow light curve \sneia found preferentially in lower SSFR and/or more massive host galaxies. Such effects must be accounted for when using \sneia as cosmological probes; for example, \citet{sullivan10} suggests the inclusion of an additional parameter in cosmological analyses to remove the host dependence.

\subsection{IMF Evolution}
By comparing the ages of the stellar population in the host galaxies with the age of the Universe at the redshift of the SNe, we can identify
the first epoch at which star-formation occurred in the host galaxies. This provides an upper limit to the formation of stars which might be
the progenitors of the Type Ia SNe. Since it is generally thought that $\lesssim8\,$M$_{\sun}$ stars are the progenitors of Type Ia SNe
we have used the host galaxy SED analysis to show that these low mass stars were in place 
3 Gyrs after the Big Bang and possibly as early as $z\sim5$ albeit with significant uncertainties that are related to the uncertainties associated
with measuring stellar population ages (Figure~\ref{fig:epochsf}). \citet{tumlinson04} argue that the nucleosynthesis yields as estimated from the metal abundances in halo stars
and the electron scattering optical depths from the cosmic microwave background
are well matched by requiring an IMF at $z\sim6$ which is truncated at $10-20$ M$_\odot$ rather than the requirement that the first stars are very massive ($>140\textrm{M}_\odot$). However, if our results are confirmed and low mass stars are found at redshifts as high as 5, this would rule out such 
a truncated IMF.

\section{Conclusion}
We have studied the host galaxies of a sample of 22 Type Ia Supernovae (SNe\,Ia) at $z \ge 0.95$ from \citet{gilliland99,blakeslee03,strolger04,riess04, riess07}. We use the broadband photometry from HST ACS \emph{BViz}, \emph{Spitzer} IRAC as well as \emph{UJHK} ground-based and some HST NICMOS \emph{JH} data from the GOODS survey. We fit the photometry to the single stellar population models of Charlot \& Bruzual (priv. com.) which are generated from the latest version of the GALAXEV code \citep{bruzual03}. We use \emph{Spitzer} MIPS 24$\mu$m data to place upper limits on the far-infrared luminosity of the hosts to break the well-known age-extinction degeneracy associated with 
SED fitting. We find the best-fit model for each host using a minimum $\chi^2$ technique to estimate the age of the stellar population of the SN\,Ia progenitors and hence place upper limits on the possible SN\,Ia delay times. We find evidence for both prompt and delayed SNe\,Ia. When the 
distribution of stellar ages in field galaxies is factored in as a control sample, we find that the prompt SNe 
are dominant over the delayed SNe and that the SN\,Ia delay times possibly have a bi-modal distribution with a paucity of SNe with delay
times of $\sim$0.8 Gyr. As a consequence of the stellar population ages, we also show that the $\lesssim8$M$_\odot$ SN\,Ia progenitor stars are in place by $z\sim2$ and possibly by $z\sim5$ (although with significant uncertainty) arguing against a truncated IMF in the first Gyr after the Big Bang. 

\acknowledgements
We are very grateful to Andy Howell, Richard Ellis and an anonymous referee for comments which improved the clarity of the paper.
We thank Chris Conselice and Rychard Bouwens for the design and reduction of the small number of NICMOS observations that were used in this paper to obtain host F160W magnitudes. We are also very grateful for the extensive telescope resources that were dedicated to the GOODS Legacy program and the contribution of various members in the analysis and reduction of those datasets. Support for this work was provided by NASA through the Spitzer Space Telescope Visiting Graduate Student Program, through a contract issued by the Jet Propulsion Laboratory, California Institute of Technology  
under a contract with NASA.

{\it Facilities:} \facility{HST (ACS, NICMOS)}, \facility{Spitzer (IRAC, MIPS)}. Several ground-based telescopes were used as part of the GOODS program: \facility{VLT (ISAAC)}, \facility{Keck}, \facility{KPNO}, \facility{CTIO}

\end{document}